\documentclass[aps,pra,shortbibliography,twocolumn,showpacs,superscriptaddress]{revtex4-1}

\usepackage{fixmath,amsmath,amsfonts,amssymb,physics,hyperref,romannum,graphicx,natbib}

\graphicspath{ {Figures/} }

\begin{document}

\newcommand{\ih}{$\textup{H}_\textup{2}^\textup{+}$}

\title{Static coherent states method for investigating high-order harmonic generation in single-electron 
molecular systems} 

\author{Mohammadreza Eidi}
\email[]{mreidi@pks.mpg.de}
\affiliation{Max Planck Institute for the Physics of Complex Systems, Dresden, Germany}

\author{Mohsen Vafaee}
\affiliation{Department of Chemistry, Tarbiat Modares University, Tehran, Iran}

\author{Hamed Koochaki Kelardeh}
\affiliation{Max Planck Institute for the Physics of Complex Systems, Dresden, Germany}

\author{Alexandra Landsman}
\affiliation{Department of Physics, The Ohio State University, Ohio, USA}

\date{\today}

\begin{abstract}
In this report, a novel methodology based on the static coherent states approach  is introduced with 
the capability of calculating various strong-field laser-induced nonlinearities in full dimensional 
single-electron molecular systems; an emphasis is made on the high-order harmonic generation. To 
evaluate the functionality of this approach, we present a case study of the Hydrogen 
molecular ion \ih interacting with a few-cycle linearly polarized optical laser with trapezoidal waveform. 
We detected that the accuracy of the obtained harmonics is considerably enhanced by averaging the 
expectation value of the acceleration of the single electron over a set of identical random simulations. 
Subsequently, the presented approach demands a significantly lower number of basis sets than the 
regular exact three dimensional unitary split-operator solvers of time dependent Schr{\"o}dinger 
equation that necessitate an extremely large number of data points in the coordinate space, and so the 
computational cost. 
Additionally, applying static coherent states method, we have investigated isolated attosecond pulse 
generation using the polarization gating technique, which combines two delayed counter rotating 
circular laser pulses, and opens up a gate at the central portion of the superposed pulse.

\end{abstract}

\pacs{}

\maketitle

\section{Introduction}
\label{sec:1}

The significant advancement of attosecond science has opened up a new field of physics where enables us to control and steer electrons using laser fields; thereby, looking into strong-field ionization and coherence dynamics of atomic, molecular and solid-state systems in their natural time scales \cite{Krausz2009,Stockman2014,Chang2016,Nisoli2017,Lin2018}. In this regard, High-order harmonic generation (HHG) from atomic and molecular and recently solid systems interacting with intense femtosecond laser pulses is widely employed as a unique and conventional tool to produce spatially and temporally coherent extreme ultraviolet \cite{Zeng2007,Popmintchev2010,Popmintchev2012}, soft and hard x-ray free-electron laser \cite{Seres2005,Ishikawa2012,Ackermann2007,Popmintchev2009} as well as isolated attosecond pulses \cite{Sansone2006,Feng2009,Goulielmakis2008,Chini2014} and attosecond pulse trains \cite{Paul2001,Winterfeldt2008}.\par

The underlying physics of the non-perturbative HHG process which is a highly nonlinear response of 
matter to the ultrashort intense laser fields can be qualitatively predicted by invoking the well-known 
semiclassical three-step model \cite{Corkum1993,Schafer1993,Krause1992a} and quantitatively by it's 
quantum version, the so-called Lewenstein model \cite{Lewenstein1994}. In these models, the electron 
which is freed to the continuum through tunneling ionization is accelerated back by the external laser 
field and a harmonic photon is emitted via recombination of the electron and its parent ion. \par

It is predicted theoretically and exhibited practically that the produced HHG spectrum which falls rapidly 
in the first few harmonics after a broad plateau ends up with a sharp cut-off. This cut-off corresponds 
to the maximum kinetic energy that electron can gain upon recombination: $K_{max} \propto 3.17U_p$ 
where 
$U_p = eE_0^2/4m\omega_0^2$ is the ponderomotive potential, $\omega_0$ is the angular frequency 
of the laser field and $E_0$ is the electric field amplitude. \par 

There are a number of approaches containing physical approximations established to compute 
nonlinear phenomena under intense ultrashort fields such as HHG spectra in atomic and molecular 
systems \cite{Baier2006,Baier2007,Chelkowski2012,Morales2014,Suarez2017}. Among them, the most 
prosperous and widely used one is the semi-analytical strong field approximation (SFA) which is 
inherent in the Lewenstein model \cite{Lewenstein1994,Lewenstein1995}. In approaches considering 
SFA, the assumption that the internal structure of the atom does not trigger HHG, oversimplifies the 
physical process. Although SFA explains high-harmonics well, it is not quite successful in describing the 
low harmonic region. Moreover, it cannot explain the behavior of the system when the Keldysh 
parameter is not small \cite{Corkum1993,Lewenstein1994}. In contrast to SFA which neglects the 
external laser field when the electron is bound to the Coulombic potential and the Coulombic potential 
when the electron is released in the continuum, the introduced approach in this article treats the 
electron-nucleus Coulombic and the laser field potentials on the same footing. For this matter we are 
obliged to accurately solve the time-dependent Schr{\"o}dinger equation (TDSE). \par

Several numerical techniques have been developed for solving the TDSE, which are either based on the 
discretization on a grid or expansions into basis functions. The most recognized approaches are: 
discrete-variable representation (DVR) \cite{Lu2008,Chen2006,Morishita2007}, finite difference 
discretization \cite{Bauer2006}, and momentum-space pseudospectral methods 
\cite{pseudospectral_Zhou2011}, finite elements and B-splines \cite{Sanz-Vicario2006,Zatsarinny2006}, 
time-dependent configuration-interaction \cite{Awasthi2005} and multiconfiguration Hartree 
\cite{Meyer2012,Caillat_Scrinzi2005}. Although the time-dependent density functional theory (TDDFT) is 
being used to approximate the quantum dynamics of some multi-electron molecular systems 
\cite{Casida2012,Castro2012}, a full understanding of the high-order harmonic generation process 
mandates full-dimensional exact solution of the TDSE. To date, such a milestone is limited to the 
single-electron systems in 3D and two-electron systems with 2D models 
\cite{Iravani2018,Suarez2017,Symonds2015,Scrinzi2012,Bandrauk2008,Guan2006,Itatani2004}.

Here, we introduce our development of static coherent state method (SCS) \cite{Eidi2018b} suited for computing high-order harmonic generation in atomic and molecular systems. SCS method solves the 3D TDSE on the base of a static grid of coherent states. Coherent states have been implementing during the last two decades as an advantageous basis set for solving 3D TDSE for high dimensional quantum systems in the presence of an ultrashort intense laser field \cite{Shalashilin2004c,Shalashilin2008a,Kirrander2011,Symonds2015,Zhou2011,Eidi2016,Eidi2018a,Eidi2018b}. One of their beneficial characteristics is that their grid can be generated randomly without forcing any boundary conditions. The major problem which other basis sets deal with is handling the Coulombic potential singularities. Coherent states simply alleviate this problem by removing the Coulombic singularities and replacing them with complex error function. \par 

Further, SCS primarily exhibits its potential in near to mid-infrared studies: The numerical complexity for solving the TDSE grows significantly with laser wavelength, and scales nearly to $\lambda ^6$ \cite{Schultz2014}. Namely, calculations at 800 nm are 64 times harder than 400 nm. This is to include only the spatial discretization. In addition, pulse duration and the temporal integration for the solution of the differential equations grow linearly with $\lambda$.\par

This is for the first time that an approach employing coherent states as its basis sets is capable of 
computing HHG in a real system solving exact 3D TDSE. Previously, HHG of a hypothetical laser-induced 
system with one electron experiencing a simple Gaussian binding potential was computed solving 1D 
TDSE employing the coupled coherent states method (CCS) \cite{Symonds2015}. CCS methods which 
are considered as trajectory-guided approaches \cite{Shalashilin2004c}, are not completely successful in 
getting high-quality convergence in real-time propagation of TDSE in single- or two-electron systems 
experiencing an external laser field \cite{Shalashilin2008a,Kirrander2011,Symonds2015}. Such issues are 
tackled in SCS by using static grids of coherent states instead of trajectory-guided ones \cite{Eidi2018b}. 
However, constructing a proper enough grid of static coherent states is challenging since we need 
more coherent states well distributed in the phase-space to cover all physical area for the simulation 
\cite{Eidi2018b}.\par

Here comes the structure of this article: we initially formulate how to compute HHG on the 
basis of the SCS method. We examine our approach by applying SCS for HHG spectra of the Hydrogen 
molecular ion \ih induced by a linearly polarized laser field and comparing it with 3D Cartesian and 
cylindrical unitary split-operator (USO) solution of TDSE \cite{Bandrauk1993,Ahmadi2014,Vafaee2004}. 
Subsequently, as a complementary work, to put SCS approach into scrutiny in a more sophisticated 
scenario, the \ih system is introduced to a ten cycle circularly polarized laser pulse with a polarization 
gate. Such a field is the central idea behind the generation of single attosecond pulses (SAP) 
\cite{Corkum1994}. \par

Throughout this paper, atomic units (a.u.), $e = \hbar = m_e = 1$ are used unless stated otherwise.
\section{Theory}
\label{sec:2}
Implementing the static coherent states method (SCS) \cite{Eidi2018b}, by representing the wave function of a single electron system as a superposition of $N$ 3D coherent states $\ket{Z}$, the time-dependent 
Schr{\"o}dinger equation 
\begin{equation}
 \label{eq:01}
 \dv{C_j}{t}=\frac{-i}{\hbar} \displaystyle\sum_{k=1}^{N} \mel**{Z_k}{H}{Z_l}D_k 
\end{equation}
 is propagated in imaginary time (ITP) until the expectation value of the field-free Hamiltonian $H$ converges to the lowest accessible value which is the ground state energy of the system for a fixed inter-nuclear distance \cite{Eidi2016}. For $C_j$ and $D_k$ coefficients in Eq. \eqref{eq:01} we have
\begin{equation}
 \label{eq:02}
 C_j= \bra{Z_j}\ket{\Psi}
\end{equation}
\begin{equation}
 \label{eq:03}
 D_k= \displaystyle\sum_{l=1}^{N} \qty(\Omega^{-1})_{kl} C_l
\end{equation} 
where $\Omega^{-1}$ is the inverse of the overlap matrix $\Omega_{kl}=\bra{Z_k}\ket{Z_l}$ with elements
\begin{equation}
 \label{eq:04}
 \Omega_{kl}=\displaystyle\sum_{j=1}^{3} 
 exp\qty(-\frac{1}{2}\qty(\abs{z_{k_j}}^2+\abs{z_{l_j}}^2)+z_{k_j}^*z_{l_j}).
\end{equation}\par
Having gained the ground state of the system in a fixed initial inter-nuclear distance $R_{12}$ at the 
end of imaginary time propagation part (ITP), we study the behaviour of the system in the presence of 
an external laser field propagating TDSE in real time (RTP). \par
To prevent nonphysical effects due to the reflection of the wave packet from the boundary, we multiply 
$C$ coefficients corresponding to each coherent state by a mask function with the form 
\cite{Krause1992b}
\begin{equation}
 \label{eq:05}
 M_{i_j}=
 \begin{cases}
		\hspace{1.5cm}1 & \abs{q_{i_j}}<Q_j		\\ \\
		cos^{\frac{1}{8}}\qty(\frac{\pi}{2}\frac{\abs{\abs{q_{i_j}}-Q_j}}{b_j}) &\abs{q_{i_j}}>Q_j
	\end{cases}
\end{equation} 
where $Q_j$ gives the boundary point in the $j^{th}$ direction, $b_j$ is the length of absorbing region 
(it can be different for each direction) and $q_{i_j}$ is the position of the $i^{th}$ coherent state in the 
$j^{th}$ direction. \par
Considering no dynamics for the nuclei, the general Hamiltonian of the system would be
\begin{equation}
 \label{eq:06}
 H=\frac{\abs{\textbf{p}_{e}}^2}{2}-\displaystyle\sum_{j=1}^{2}\frac{1}{\abs{\textbf{r}_{e}-\textbf{R}_{j}}}+\frac{1}{\abs{\textbf{R}_1-\textbf{R}_2}}+\textbf{r}_{e}.\textbf{E}(t).
\end{equation} 
In Eqs. \eqref{eq:06}, the first term is for the kinetic energy of the single electron, the second term stands for the electron-nuclear Coulombic potentials and the third term is the constant repulsive potential from the two nuclei. In the RTP part of the simulation, considering the dipole approximation in the length gauge, the forth term is due to the presence of an external laser field. \par
The matrix elements of the kinetic energy of the single electron in Eq. \eqref{eq:06} on the base of a 3D CS grid could be achieved by \cite{Eidi2016} 
\begin{equation}
 \label{eq:07}
 \footnotesize{\mel**{Z_k}{\frac{\abs{\textbf{p}_{e}}^2}{2}}{Z_l}=-\frac{\gamma}{2}\bra{Z_k}\ket{Z_l}\displaystyle\sum_{j=1}^{3}\qty({z_k^*}_j^2+{z_l}_j^2-2{{z_k}_j^*}{{z_l}_j}-1)}
\end{equation} 
where $j$ is the dimension number.
For the matrix elements of electron-nuclear Coulombic potentials in Eq. \eqref{eq:06} one can derive 
\begin{equation}
 \label{eq:08}
 \mel**{Z_k}{\frac{1}{\abs{\textbf{r}_{e}-\textbf{R}_{i}}}}{Z_l}=\bra{Z_k}\ket{Z_l}\frac{1}{\sqrt{\abs{\mathbold{\rho}_{e_i}}^2}}erf\qty(\sqrt{\gamma\abs{\mathbold{\rho}_{e_i}}^2})
\end{equation}
where $i$ is the index number of the nuclei and 
\begin{equation}
 \label{eq:09}
	\mathbold{\rho}_{e_i}=\frac{Z_{k}^*+Z_{l}}{\sqrt{2\gamma}}-\textbf{R}_{i}. 
\end{equation}\par
Here, we compute the HHG spectrum $D(\omega)$ as the squared magnitude of the Fourier transforms (FT) of the expectation value of the electron dipole acceleration $\textbf{a}_{e}$
\begin{equation}
\label{eq:10}
D(\omega)=\abs{\int_{0}^{T}{\mel**{\psi}{\textbf{a}_{e}}{\psi} H(t)e^{-i\omega{t}}dt}}^2
\end{equation} 
where $T$ is the total pulse duration and
\begin{equation}
\label{eq:11}
H(t)=\frac{1}{2}\qty[1-cos\qty(2\pi\frac{t}{T})]
\end{equation}
is the Hanning function which filters nonphysical features (non decaying components) from the HHG spectrum as the Fourier transform is applied over a finite time. \par
The expectation value of the dipole acceleration of the single electron in \ih system can be computed using the Newtonian equation of motion
\begin{equation}
 \label{eq:12}
 \mel**{\psi}{\textbf{a}_{e}}{\psi}=\frac{1}{{m_e}}\mel**{\psi}{\textbf{F}_{1e}+\textbf{F}_{2e}+\textbf{F}_{le}}{\psi}.
\end{equation} 
where $\textbf{F}_{1e}$ and $\textbf{F}_{2e}$ are the expectation value of the nucleus-electron attractive forces, $\textbf{F}_{le}$ is the force exerted on the electron by the external laser field and $m_e$ is the mass of electron.\par
For the expectation value of the nucleus-electron attractive forces one can verify that \cite{Eidi2018b}
\begin{equation}
 \label{eq:13}
 \mel**{\psi}{\textbf{F}_{ie}}{\psi} =\displaystyle\sum_{kl} \textbf{F}_{ie_{kl}} D_k^* D_l
\end{equation} 
where 
\begin{equation}
 \label{eq:14}
 \textbf{F}_{ie_{kl}}=\mel**{Z_k}{\frac{-\textbf{r}_{ei}}{\abs{\textbf{r}_{ei}}^3}}{Z_l} 
\end{equation} 
is the matrix elements of the attractive Coulombic force on the base of a static grid of coherent states and $\textbf{r}_{ei}=\textbf{r}_e-\textbf{R}_i$ . To compute $\textbf{F}_{ie_{kl}}$, applying the identity operator of coordinate states of electrons leads to
\begin{equation}
 \label{eq:15}
\footnotesize{\textbf{F}_{ie_{kl}}=-\int_{-\infty}^\infty \int_{-\infty}^\infty \bra{Z_k}\ket{\textbf{r}_e}\mel**{\textbf{r}_e}{\frac{\textbf{r}_{ei}}{\abs{\textbf{r}_{ei}}^3}}{\textbf{r}_e^\prime}\bra{\textbf{r}_e^\prime}\ket{Z_l}d{\textbf{r}_e}d{\textbf{r}_e^\prime}}.
\end{equation} 
Employing the continuous Dirac delta function in the coordinate representation 
\begin{equation}
\label{eq:16}
\begin{cases}
\mel**{\textbf{r}_e}{f(\textbf{r}_e)}{\textbf{r}_e^\prime}=\delta(\textbf{r}_e-\textbf{r}_e^\prime)f(\textbf{r}_e)\\ \\
\int_{-\infty}^\infty f(\textbf{r}_e)\delta(\textbf{r}_e-\textbf{r}_e^\prime)d{\textbf{r}_e}=f(\textbf{r}_e^\prime)
\end{cases}
\end{equation} 
\\
one gets
\begin{equation}
\label{eq:17}
\textbf{F}_{ie_{kl}}=-\int_{-\infty}^\infty \bra{Z_k}\ket{\textbf{r}_e}\bra{\textbf{r}_e}\ket{Z_l}\frac{\textbf{r}_{ei}}{\abs{\textbf{r}_{ei}}^3}d{\textbf{r}_e}.
\end{equation} 
Using the fact that coherent states are Gaussian wave packets in the coordinate representation
\begin{equation}
 \label{eq:18}
\bra{Z_k}\ket{\textbf{r}_e}=(\frac{\gamma}{\pi})^{3/4}e^\qty(\frac{-\gamma}{2}(\textbf{r}_e-\frac{\sqrt{2}Z_k^*}{\gamma^{\frac{1}{2}}})^2+(\frac{Z_k^*-Z_k}{2})Z_k^*).
\end{equation} 
and by applying the Gaussian product rule \cite{Helgaker1995}, it could be verified that 
\begin{equation}
\label{eq:19}
\bra{Z_k}\ket{\textbf{r}_e}\bra{\textbf{r}_e}\ket{Z_l}=(\frac{\gamma}{\pi})^{3/2}\bra{Z_k}\ket{Z_l}e^{-\gamma\abs{\textbf{r}_{c_1}}^2}
\end{equation} 
where
\begin{equation}
\label{eq:20}
\textbf{r}_{c_1}=\textbf{r}_e-\textbf{c}_1\hspace{0.3cm},\hspace{0.3cm}\textbf{c}_1=\frac{Z_k^*+Z_l}{\sqrt{2\gamma}}.
\end{equation} 
Taking into account the over-completeness property of coherent states from Eq. \eqref{eq:04} and then substituting Eq. \eqref{eq:19} into Eq. \eqref{eq:17} one can easily get
\begin{equation}
\label{eq:21}
\textbf{F}_{ie_{kl}}=-\bra{Z_k}\ket{Z_l}(\frac{\gamma}{\pi})^{3/2}\int_{-\infty}^\infty\frac{\textbf{r}_{ei}}{{\abs{\textbf{r}_{ei}}}^3}e^{-\gamma\abs{\textbf{r}_{c_1}}^2}d\textbf{r}_e.
\end{equation} 
Now substituting this Laplace transform
\begin{equation}
\label{eq:22}
\frac{1}{\abs{\textbf{r}_{ei}}^3}=\frac{4}{\sqrt{3}\pi}\int_0^\infty e^\qty(t^{-2/3}\abs{\textbf{r}_{ei}}^2)dt
\end{equation}
into Eq. \eqref{eq:21} and applying again the Gaussian product rule leads to \\
\begin{equation}
\label{eq:23}
\begin{array}{l}
\textbf{F}_{ie_{kl}}=-\bra{Z_k}\ket{Z_l}\frac{4{\gamma^{3/2}}}{3\pi^2} \int_0^\infty 
e^\qty(-\frac{\gamma{t^{2/3}}}{\gamma+t^{2/3}}\abs{\mathbold{\rho}_{ei}}^2)\\ 
\hspace{1.4cm}\int_{-\infty}^\infty\textbf{r}_{ei} 
e^\qty(-\qty(\gamma+t^{2/3})\abs{\textbf{r}_{{c_2}_i}}^2)d{\textbf{r}_i}dt
\end{array}
\end{equation} 
where
\begin{equation}
\label{eq:24}
\mathbold{\rho}_{ei}=\frac{Z_k^*+Z_l}{\sqrt{2\gamma}}-\textbf{R}_i
\end{equation} 
\begin{equation}
\label{eq:25}
\textbf{r}_{{c_2}_i}=\textbf{r}_i-{\textbf{c}_2}_i\hspace{0.3cm},\hspace{0.3cm}{\textbf{c}_2}_i=\frac{\gamma}{\gamma+t^{2/3}}{\textbf{c}_1}+\frac{t^{2/3}}{\gamma+t^{2/3}}\textbf{R}_i.
\end{equation}
One can also show that 
\begin{equation}
\label{eq:26}
\textbf{r}_{ei}=\textbf{r}_{{c_2}_i}+\frac{\gamma}{\gamma+t^{2/3}}\mathbold{\rho}_{ei}.
\end{equation} 
Substituting Eq. \eqref{eq:26} in Eq. \eqref{eq:23} and applying the well-known 3D Gaussian integral
\begin{equation}
\label{eq:27}
\int_{-\infty}^\infty e^{-\alpha\textbf{r}^2}d\textbf{r}=\qty(\frac{\pi}{\alpha})^{3/2}
\end{equation} 
and considering
\begin{equation}
\label{eq:28}
\frac{t^{2/3}}{\gamma+t^{2/3}}=u^2
\end{equation}
it is straightforward to verify that 
\begin{equation}
\label{eq:29}
\textbf{F}_{ie_{kl}}=-\qty(\frac{4\gamma^3}{\pi})^{1/2}\mathbold{\rho}_{ei} B_1\qty(\gamma\abs{\mathbold{\rho}_{ei}}^2)\bra{Z_k}\ket{Z_l}
\end{equation} 
where $B_1$ is the first order Boys function
\begin{equation}
\label{eq:30}
B_1(x)=\int_0^1 t^2 e^{-xt^2}dt.
\end{equation} \par
For the electric force exerted on each nucleus by the external laser field ($\textbf{E}$) 
we also simply get 
\begin{equation}
\label{eq:31}
 \mel**{\psi}{\textbf{F}_{le}}{\psi}=-e\textbf{E}.
\end{equation} 

\section{Results and Discussion}
\label{sec:3}
We now turn to employ our proposed SCS approach and investigate HHG in \ih for different laser induced scenarios. We initially study HHG from a linearly polarized optical field, and afterward using polarization gating technique explore the single isolated attosecond pulse generation in \ih. It is worth mentioning that since \ih is an oriented heteronuclear molecule, the spatial symmetry is broken in the system and both odd and even harmonics are allowed. \par
In these applications of the SCS method, we utilize two complementary CS grid boxes with $N_i$ CS in 
the internal box and $N_e$ CS in the external box which form a complex static CS grid with a total 
number of $N_t=N_i+N_e$ coherent states \cite{Eidi2018b}. Coherent states of the external box which 
are distributed differently compared to internal ones, play a critical stabilizing role in the real time 
propagation of TDSE in the presence of an external laser field. 

\subsection{HHG by linear laser field} \label{subsec:linear}

At first, the behavior of \ih system is studied at $R_{12}=2$ a.u. and $R_{12}=3$ a.u. in the presence of a 
trapezoidal 800 nm 5-cycles (which rises in the first cycle and falls in the last one) linear laser field with 
intensity of $I=10^{14}$ W/cm$^2$ to examine the effectiveness and performance of SCS in such kind of 
simulations. The external laser field is linearly polarized along the z axis with the following shape
\begin{equation}
\label{eq:32}
\textbf{E}(t)=A_{env}(t)E_0 cos(\omega t)\hat{k}
\end{equation} 
where $E_0$ is the amplitude of the laser field, $\omega$ is the carrier frequency and 
$A_{env}(t)$ is the trapezoidal envelope function. \par

\begin{figure} 
	\includegraphics[width=0.5\textwidth]{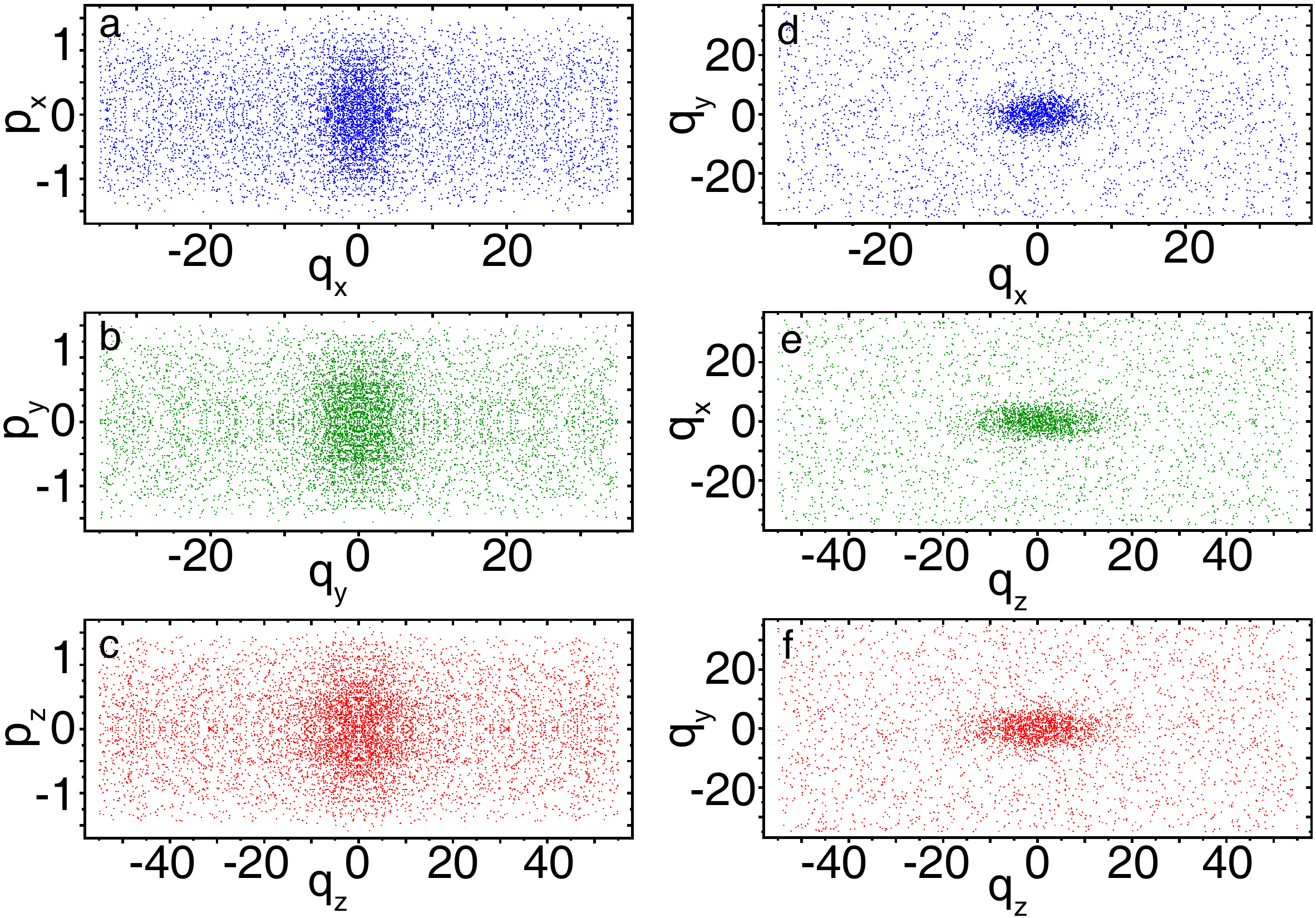}
	\caption{\label{Fig.1-QP-L}[Color online] Phase (left panel) and coordinate space (right panel) representations of a random 3D grid for simulating HHG in \ih induced by a linearly polarized laser field in the z direction. Hence, the external coherent states are distributed more widely along the z direction compared to x and y. A total number of 8000 CS implemented in each simulation, containing 3000 CS distributed around the two nuclei and 5000 CS elsewhere to capture phenomena occurring beyond the effective Coulombic range.}
\end{figure} 

\begin{figure*} 
	\includegraphics[width=1.0\textwidth]{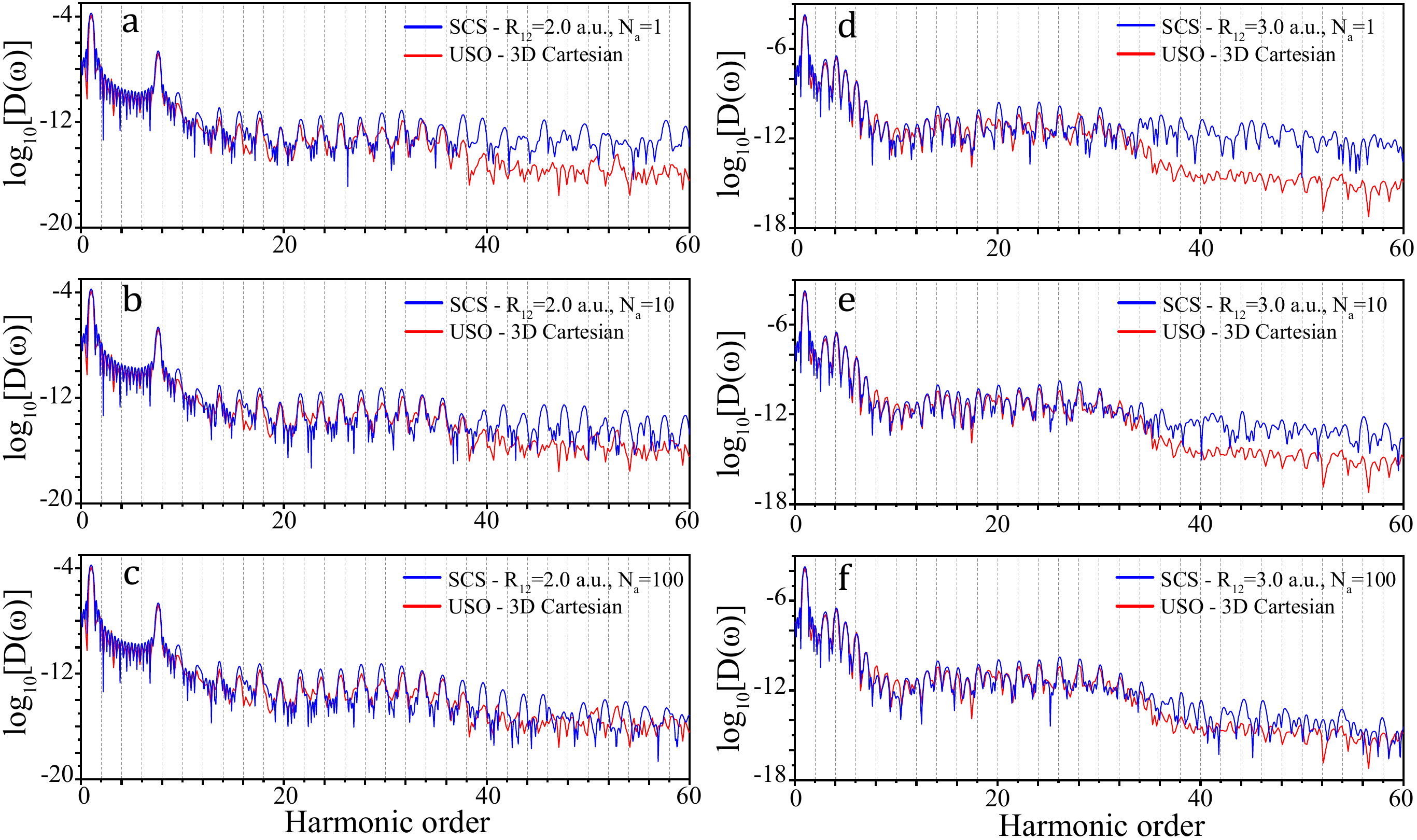}
	\caption{\label{Fig.2-HHG-800}[Color online] The HHG spectrum of \ih calculated based on the SCS at 
		$R_{12}=2.0$ a.u. (left panel) and $R_{12}=3.0$ a.u. (right panel), induced by a 5-cycles trapezoidal 
		800 nm laser field with the intensity of $I=10^{14}$ W/cm$^2$ in comparison with the 3D Cartesian 
		USO solution of TDSE under the same conditions (i.e., laser parameters and grid boundaries). For 
		both internuclear distances $R_{12}$, three distinct cases with averaging factor $N_a$= 1, 10 and 
		100 are considered. We improve the observed deviation between our results and the Cartesian 
		USO solver in the above cut-off region, as we increase the number of averaged random 
		simulations 
		$N_a$. } 
\end{figure*} \par

For simulating HHG in \ih at $R_{12}=2$ a.u. and $R_{12}=3$ a.u. introduced to the mentioned laser 
field, a complex static grid with a total number of 8000 CS is constructed distributing 3000 randomly 
generated CS by the Gaussian distribution function in the internal box around the two nuclei and 
distributing 5000 randomly generated CS more homogeneously in the external box by using the Sobol 
sequence \cite{Joe2003}. The nuclei are considered to be located in the z direction. The external 
coherent states are randomly distributed in the phase space between (-35, 35)  a.u. in x, y , (-55, 
55)  a.u. in z direction and (-2.0, 2.0)  a.u. in all momentum directions. The gamma parameter tunes 
the width of coherent states in phase space and is considered to be $\gamma=0.7$ for all coherent 
states \cite{Eidi2016}. The compression parameters \cite{Eidi2016} in $\qty{x, y, z, p_x, p_y, p_z}$ 
directions are equal to $\qty{0.5, 0.5, 0.25, 2.0, 2.0, 2.0}$ and $\qty{1.0, 1.0, 1.0, 1.0, 1.0, 1.0}$, 
respectively for the internal and external boxes. Absorbing boundaries ($Q$) in x, y and z directions are 
also considered to be 30, 30 and 50 a.u., respectively. The coordinate and phase space representation 
of one of such random 3D coherent state grids is depicted in FIG. \ref{Fig.1-QP-L}. As depicted, the 
region in proximity of the nuclei should involve more CSs to capture short-range phenomena such as 
Coulombic potential effects. Constructed grid size is large enough to cover the predicted 
ponderomotive radius $\alpha_p=\frac{E_0}{\omega^2}$. Considering the intensity of $I=10^{14}$ 
W/cm$^2$ for the external laser field for this part of our study, $ \alpha_p $ would be $16.45$ a.u. \par 

In Fig. \ref{Fig.2-HHG-800}, we have respectively depicted the resulted HHG spectrum for $R_{12}=2$ 
a.u. (left panel) and $R_{12}=3$ a.u. (right panel) from SCS approach and compared it to the results of 
3D Cartesian USO solution of TDSE with the same conditions. We realized an imperative 
characteristics of the proposed SCS-based method which notably improves the results  at a relatively 
small computational cost; a significantly more CSs can be incorporated by repeating the simulation 
based on a number of similar random CS 
grids and correspondingly averaging the expectation value of the acceleration of an electron along the 
z direction. The averaging 
factor, $N_a$, defines the number of recurring random simulations. As can be seen in Fig. 
\ref{Fig.2-HHG-800}a and Fig. \ref{Fig.2-HHG-800}d, even 
with only 8000 CS ($N_a=1$) the spectrum obtained from SCS is in agreement with that of 3D Cartesian 
USO in the low harmonic regions. In addition, the consistency of SCS with 3D Cartesian USO is 
acceptable in the plateau region (for harmonic orders lower than 36). The inconsistency which arises 
after cut-off (for harmonic orders higher than 36), can be alleviated including more CS into the 
simulation, i.e. increasing $N_a$ (cf. Fig.\ref{Fig.2-HHG-800}). 
 Incorporating a high number of CS into one simulation, on the other hand, is either not practically 
 feasible or is exceedingly expensive. As it is evident form Fig. \ref{Fig.2-HHG-800}(b-c) and Fig. 
 \ref{Fig.2-HHG-800}(e-f), increasing $N_a$ and hence including more CS into the simulation (up to 
 800,000 for $N_a$=100), the results show a better cohesion with those of 3D Cartesian USO method. 
From Fig \ref{Fig.2-HHG-800}(a-f) it can be conceived that for higher internuclear distances the obtained 
HHG spectrum is more intense. However, as the internuclear distance is increased, more deviation 
from 3D Cartesian USO solver is seen in the above cut-off region. This issue arises from the fact that at 
higher nuclear distances the electronic cloud becomes broader and correspondingly, using CS 
grids with an identical and adequate box size, a higher number of coherent states are demanded in 
order to capture HHG at higher internuclear distances $R_{12}$. 
 \par

\begin{figure} 
	\includegraphics[width=0.50\textwidth]{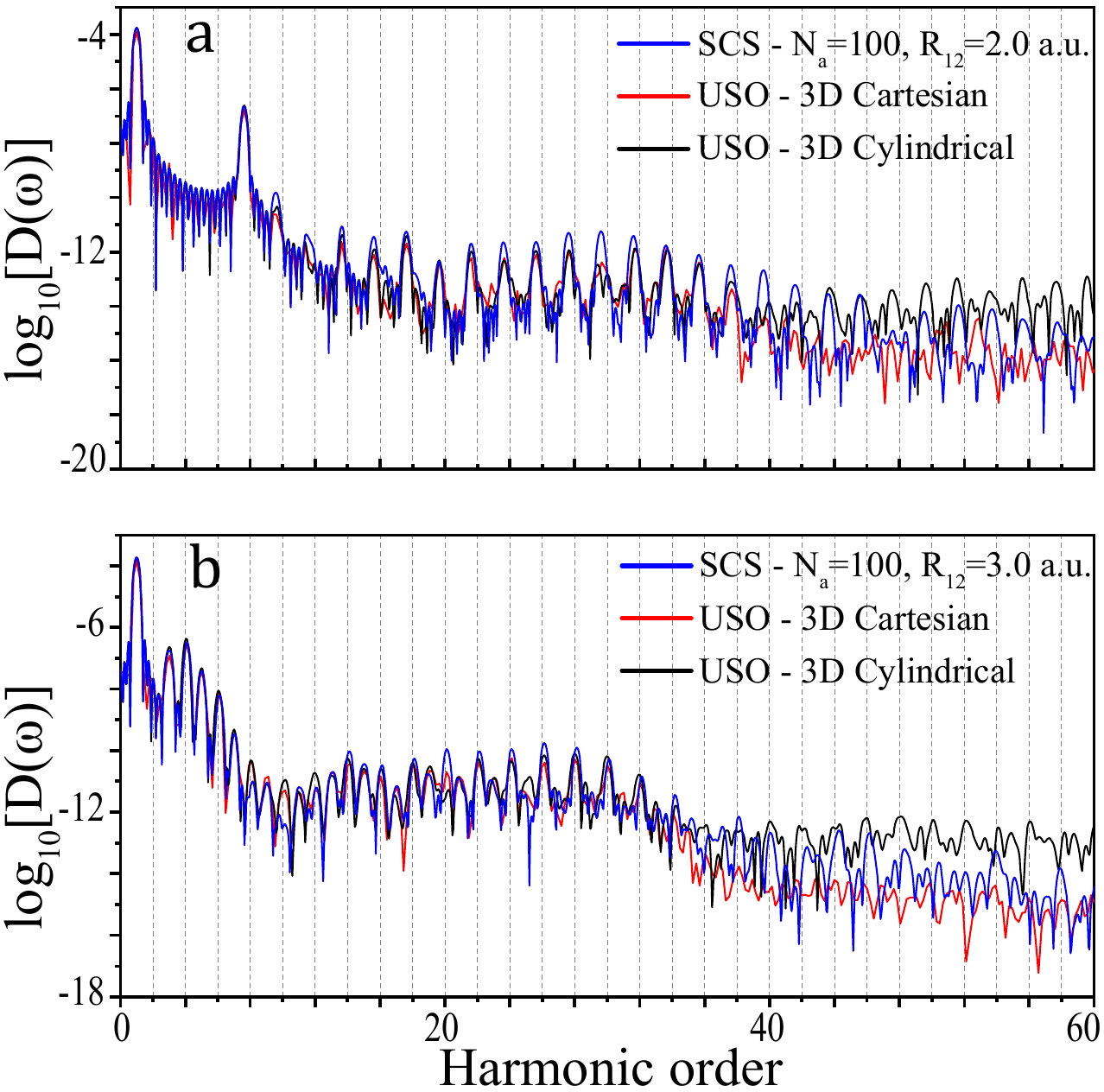}
	\caption{\label{Fig.3-SCS-Rho-xyz}[Color online] HHG spectrum of \ih at(a) $R_{12}=2$ a.u. and (b) 
	$R_{12}=3$ a.u., obtained from SCS method averaging over $N_a=100$ random simulations in 
	comparison with 3D Cartesian and cylindrical USO TDSE solvers under identical conditions. The laser 
	parameters are the same as Fig. \ref{Fig.2-HHG-800}. The HHG spectrum from SCS above the cut-off 
	is more consistent with the Cartesian USO solver than the Cylindrical one. } 
\end{figure} \par

For the 3D Cartesian USO solution of TDSE the grid size is considered to be almost the same as those 
we used in SCS. However, a total number of 42,250,000 grid points (324, 324 and 400 grid points in x, y 
and z directions, respectively) are taken into account. Using such a high number of grid points to 
compute HHG spectrum of the system is not computationally cost-effective. One can instead exploit the 
3D cylindrical USO solution of TDSE \cite{Ahmadi2014}, which due to its intrinsic cylindrical symmetry is 
less costly and requires fewer grid points (a total number of 259,200) compared to 3D Cartesian USO. In 
Fig.\ref{Fig.3-SCS-Rho-xyz}, we evaluate the SCS result with ($N_a$= 100) for $R_{12}=2$ a.u. and 
$R_{12}=3$ a.u. in comparison with both 3D Cartesian and cylindrical USO results.
As can be concluded from Fig. \ref{Fig.3-SCS-Rho-xyz}, in low harmonic regions and up to the cut-off, all 
results are almost in agreement with each other. For higher harmonics, the SCS results are closer to 
the 3D Cartesian USO. Comparison of SCS results with those of 3D Cartesian and cylindrical USO 
exhibits that even the two compared USO solutions do not illustrate a full cohesion. 
However, due to the fact that the harmonic spectrum above the cut-off region is highly sensitive to the 
parametrization of the system and the used frames, the full consistency between results of different 
approaches might be unattainable and it is reasonable to expect deviation in the calculated emission 
spectrum from different methods. 

\subsection{Single attosecond pulse generation } 
\label{subsec:SAP}
For the next step, to evaluate SCS approach in simulation of more complex laser-induced scenarios, the single isolated attosecond pulse (SAP) generation is investigated in \ih using polarization gating technique \cite{Corkum1994}. Such a polarization gate is generated without spatial filtering in central part of the pulse by superposing two left ($-$) and right-hand ($+$) circularly polarized Gaussian pulses propagated in the z direction
\begin{equation}
\label{eq:33}
\footnotesize{E_{\pm}(t) =E_0 \hspace{0.1cm} e^{-2\ln{2}((t - {t_d}/2)/{\tau_p})^2}\qty(\cos(\omega 
t+\phi) \hat{x} \pm \sin (\omega t+\phi) \hat{y})}
\end{equation}
in which $E_0$, $t_d$, $\tau_p$, $\omega$ and $\phi$ are the field amplitude, time delay between two pulses, the full width at half maximum (FWHM) of the the Gaussian shaped pulse, the carrier frequency and the carrier-envelope phase, respectively. \par
Having computed x- or y-component of dipole acceleration $a_{x(y)}$ via Eq. \eqref{eq:12}, the profile of the attosecond pulse for each direction is obtainable superposing different harmonics orders \cite{Safaei2018}
\begin{equation}
\label{eq:34}
I_{x(y)}(t) = \abs{\displaystyle\sum a_{{x(y)}_q}e^{iq\omega t}}^2\\
\end{equation}
where
\begin{equation}
\label{eq:35}
a_{{x(y)}_q}= \int{a_{x(y)}(t) e^{-iqwt}\rm{d}t}.
\end{equation}
The time-frequency profiles of the high harmonics $w(\omega ,t)$ are also obtained via a Morlet wavelet transform of the time-dependent dipole acceleration 
\begin{equation}
\label{eq:36}
w(\omega ,t) = \sqrt {\frac{\omega }{\pi ^{\frac{1}{2}}\sigma }}\int_{ - \infty }^\infty {a} ({t^\prime }) e^{- 
i\omega (t^\prime-t)} e^{- \frac{{\omega ^2}{(t^\prime-t)^2}}{2{\sigma 
^2}}}d{t^\prime }
\end{equation}
where we set the Gaussian width $ \sigma=2\pi $ in this work.\par

In order to study how SAP is generated in \ih at $R_{12}=3$ a.u. using SCS, the system is simulated in 
the presence of a 800 nm 10-cycle laser field which has a polarization gating in the middle of the pulse. 
Such laser field is formed by combining two 8-cycle left and right-hand circularly polarized Gaussian 
pulses with both time delay $t_d$ and FWHM of the Gaussian envelope $\tau_p$ equal to 2 cycles (i.e., 
5.33702 fs or 220.64 a.u.) and the intensity of $ I=3\times 10^{14}$ W/cm$^2$ for $\phi=0$ and 
$\phi=\pi/2$ similar to \cite{Safaei2018}.

\begin{figure} 
	\includegraphics[width=0.5\textwidth]{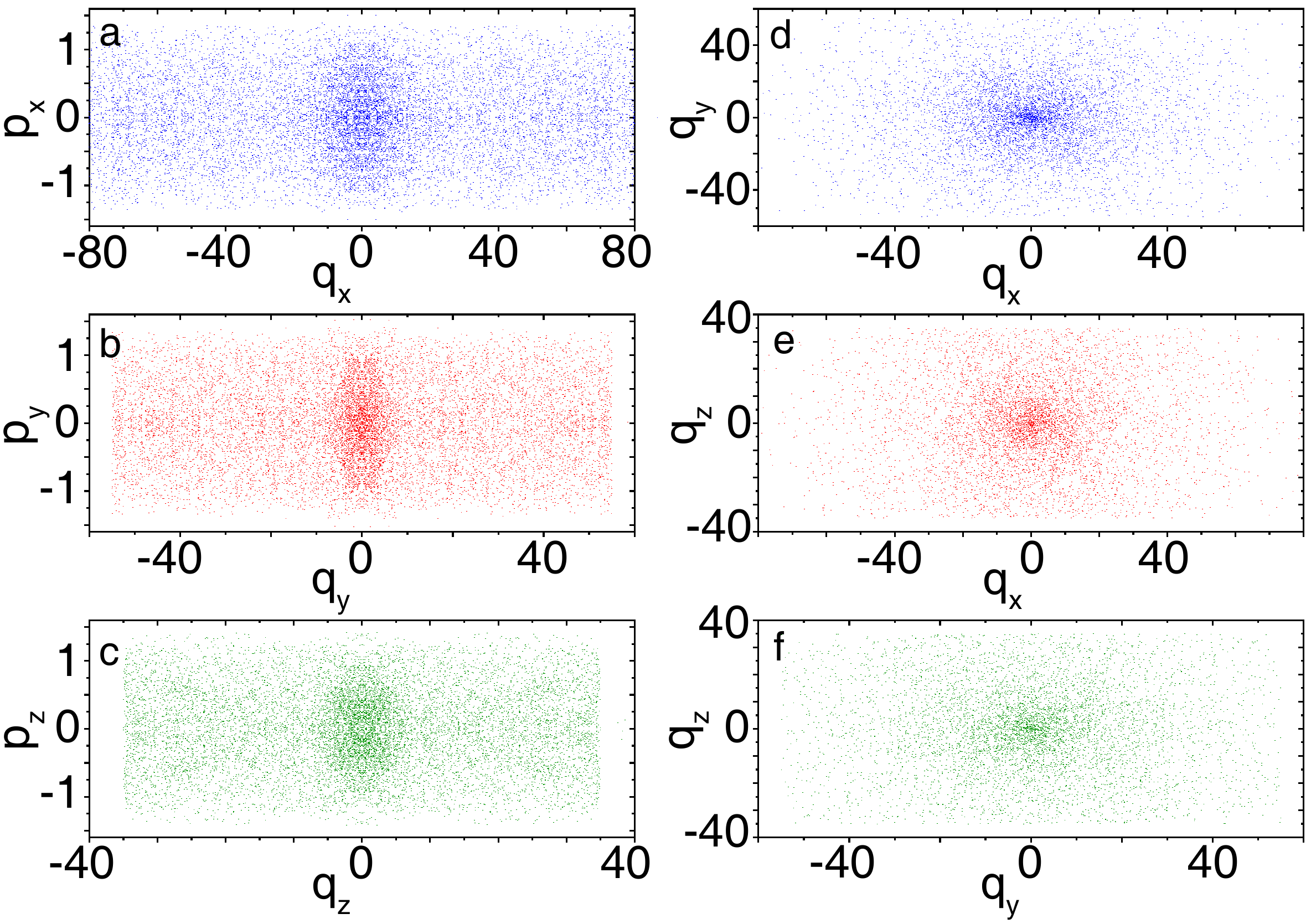}
	\caption{\label{Fig.4-QP-Atto}[Color online] 
		Phase (left panel) and coordinate space (right panel) representations of a random 3D grid for 
		investigating the generation of the single attosecond pulse scenario in \ih at $R_{12}=3.0$ a.u. 
		using the polarization gating technique \cite{Corkum1994}. A total number of 12000 coherent 
		states is taken in each simulation, with 3000 CS distributed around the two nuclei and 9000 CS in 
		the farther region to capture phenomena occurring beyond the effective Coulombic range. Since 
		at the polarization gate the field is mostly polarized along the x direction, the external coherent 
		states are distributed more widely along this direction compared to y and z directions. }
\end{figure} \par

\begin{figure*} 
	\includegraphics[width=1.0\textwidth]{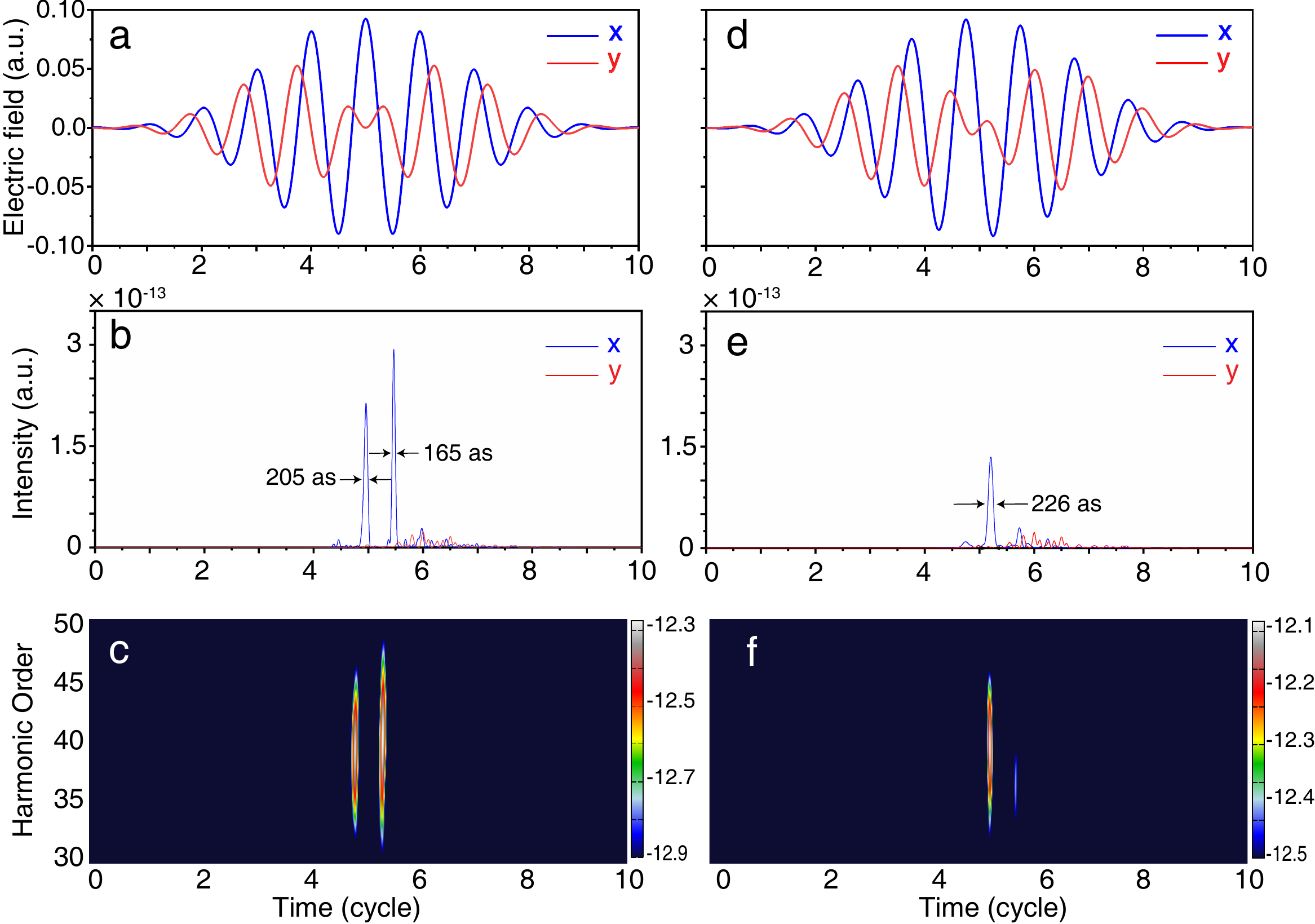}
	\caption{\label{Fig.5-AttoL}[Color online] Single isolated attosecond pulse generation. Left panel (a-c) 
	for carrier envelop phase $\phi=0$: and right panel (d-f) for carrier envelop phase $\phi=\frac{\pi}{2}$. 
	The first row (a,d) plots x (blue) and y (red) components of 10-cycle electric field constructed by 
	combining two 8-cycle left and right-hand circularly polarized Gaussian pulses with both time delay 
	$t_d$ and FWHM $\tau_p$ equal to 2 cycles and $I=3\times 10^{14}$ W/cm$^2$. The second row (c,f) 
	depicts isolated attosecond pulse from \ih in x (blue) and y (red) directions. Since at the gate both 
	fields are primarily polarized in the x direction, the generated pulse amplitudes in y direction are 
	significantly weaker than their corresponding x counterparts. The last row (b,e) illustrates the 
	corresponding Morlet wavelet time-frequency profiles in the x direction.}
\end{figure*} 

To investigate SAP in \ih introduced to such laser field, a complex static grid with a total number of 
12000 CS is constructed distributing 3000 randomly generated CS by the Gaussian distribution function 
in the internal box around the two nuclei and distributing 9000 randomly generated CS more 
homogeneously in the external box by using the Sobol sequence. The nuclei are considered to be 
located in the x direction. The external coherent states are randomly distributed in the phase space 
between (-80, 80)  a.u. in x direction, (-55, 55) a.u. in y direction, (-35, 35) a.u. in z direction 
and (-2.0, 2.0)  a.u. in all momentum directions. The gamma parameter for all coherent states is 
$\gamma=0.7$. The compression parameters for the internal and the external boxes in $\qty{x, y, z, 
p_x, p_y, p_z}$ directions are respectively equal to $\qty{0.25, 0.5, 0.5, 2.0, 2.0, 2.0}$ and $\qty{1.0, 1.0, 
1.0, 1.0, 1.0, 1.0}$. Absorbing boundaries ($Q$) in x, y and z directions are also considered to be 75, 50 
and 30 a.u., respectively. Coordinate and phase space representation of one of such random 3D 
coherent state grids is delineated in FIG. \ref{Fig.4-QP-Atto}.\par 

In Fig. \ref{Fig.5-AttoL}, we have illustrated the SAP achieved from SCS approach using polarization 
gating technique for carrier envelop phase $\phi=0$ (left panel) and $\phi=\pi/2$ (right panel). The 
simulation results in Fig. \ref{Fig.5-AttoL} obtained by averaging the expectation value of the 
acceleration of the single electron along x and y directions over 40 different random simulations (each 
contains a total number of 12000 CS). In the first row of Fig. \ref{Fig.5-AttoL} we have plotted x and y 
components of 10 cycles laser pulses for carrier envelope phases $\phi=0$ (left) and $\phi=\pi/2$ (right). 
In the second row of Fig. \ref{Fig.5-AttoL} we have depicted the profile of SAP pulse generated from 
high-order harmonic spectrum of \ih for carrier envelope phases $\phi=0$ (left) and $\phi=\pi/2$ (right) 
both in x and y directions. Depicted x and y components of SAP are created by superposing the 
harmonics of plateau from the 30th to the 50th orders as it is shown in the third row of Fig. 
\ref{Fig.5-AttoL} which delineates the corresponding Morlet wavelet time profiles in the x direction.

As it is apparent, at the central part of both fields (Fig. \ref{Fig.5-AttoL}a and Fig. \ref{Fig.5-AttoL}d)) which are defined as polarization gate, the y component $E_y(t)$ is suppressed and consequently ellipticity of both pulses is changed from circular to linear. Therefore the x component of the pulse $E_x(t)$ becomes the main driving field and hence at the central portion of both pulses isolated attosecond pulse is generated. This 
conclusion is in agreement with extremely low intensity taps generated in the y direction for both $\phi=0$ (Fig. \ref{Fig.5-AttoL}b) and $\phi=\pi/2$ (Fig. \ref{Fig.5-AttoL}e). In the x direction, for the laser pulse with $\phi=0$, two pulses are generated with a comparable intensity (with duration of 165 as and 205 as), while for the case of $\phi=\pi/2$, only one pulse with a comparable intensity (with duration of 226 as) is left. 

\section{Conclusions}
\label{sec:4}

In summary, a static coherent states (SCS) method is developed for full-dimensional quantum simulation of high-order harmonic generation process in single-electron systems such as Hydrogen molecular ion \ih. Such an approach using coherent state as the basis sets of the electronic system, for the first time, successfully calculates the high-order harmonic generation (HHG) spectrum of a full-dimensional realistic system. 
A key significance of SCS approach is that the choice of basis sets possess randomness attributes. As a 
result, the correctness of our simulation can be increased by averaging over calculations from the 
ensemble of random CS grids.
As the first case study, we have examined the correctness and optimality of our full dimensional simulation of HHG for \ih in the presence of a linearly polarized laser field, comparing them with 3D unitary split-operator (USO) solvers. Compared to the USO solutions, SCS exhibits a higher cost effectiveness for calculating the HHG spectrum. \par
In the second scenario, as a complementary examination of the SCS approach, the generation of single attosecond pulse (SAP) is performed using the polarization gating technique. Such a polarization gate is constructed by mixing two delayed Gaussian pulses with opposite circularity propagating in the z direction, in which the gate is opened up at the central portion of the generated pulse. At the polarization gate, we show that the field polarization dominantly turns into linear in the x direction. Consequently, isolated attosecond pulses generated in the x direction are considerably more intense than those generated in the y direction. \par
Presented approach in this work demands a considerably lower number of coherent states in the phase space compared to the USO solvers which requires exceedingly large number of grid points in the coordinate space. Such a characteristics shows its potent in computing higher dimensional systems such as $\textup{H}_\textup{2}$, where obtaining the exact solutions is limited by the available methods. The detailed analysis associated with such systems is currently under development.

\def\bibsection{\section*{REFERENCES}}
\bibliographystyle{apsrev4-1}
\bibliography{References}

\begin{thebibliography}{62}%
\makeatletter
\providecommand \@ifxundefined [1]{%
 \@ifx{#1\undefined}
}%
\providecommand \@ifnum [1]{%
 \ifnum #1\expandafter \@firstoftwo
 \else \expandafter \@secondoftwo
 \fi
}%
\providecommand \@ifx [1]{%
 \ifx #1\expandafter \@firstoftwo
 \else \expandafter \@secondoftwo
 \fi
}%
\providecommand \natexlab [1]{#1}%
\providecommand \enquote  [1]{``#1''}%
\providecommand \bibnamefont  [1]{#1}%
\providecommand \bibfnamefont [1]{#1}%
\providecommand \citenamefont [1]{#1}%
\providecommand \href@noop [0]{\@secondoftwo}%
\providecommand \href [0]{\begingroup \@sanitize@url \@href}%
\providecommand \@href[1]{\@@startlink{#1}\@@href}%
\providecommand \@@href[1]{\endgroup#1\@@endlink}%
\providecommand \@sanitize@url [0]{\catcode `\\12\catcode `\$12\catcode
  `\&12\catcode `\#12\catcode `\^12\catcode `\_12\catcode `\%12\relax}%
\providecommand \@@startlink[1]{}%
\providecommand \@@endlink[0]{}%
\providecommand \url  [0]{\begingroup\@sanitize@url \@url }%
\providecommand \@url [1]{\endgroup\@href {#1}{\urlprefix }}%
\providecommand \urlprefix  [0]{URL }%
\providecommand \Eprint [0]{\href }%
\providecommand \doibase [0]{http://dx.doi.org/}%
\providecommand \selectlanguage [0]{\@gobble}%
\providecommand \bibinfo  [0]{\@secondoftwo}%
\providecommand \bibfield  [0]{\@secondoftwo}%
\providecommand \translation [1]{[#1]}%
\providecommand \BibitemOpen [0]{}%
\providecommand \bibitemStop [0]{}%
\providecommand \bibitemNoStop [0]{.\EOS\space}%
\providecommand \EOS [0]{\spacefactor3000\relax}%
\providecommand \BibitemShut  [1]{\csname bibitem#1\endcsname}%
\let\auto@bib@innerbib\@empty
\bibitem [{\citenamefont {Krausz}\ and\ \citenamefont
  {Ivanov}(2009)}]{Krausz2009}%
  \BibitemOpen
  \bibfield  {author} {\bibinfo {author} {\bibfnamefont {F.}~\bibnamefont
  {Krausz}}\ and\ \bibinfo {author} {\bibfnamefont {M.}~\bibnamefont
  {Ivanov}},\ }\href {\doibase 10.1103/RevModPhys.81.163} {\bibfield  {journal}
  {\bibinfo  {journal} {Reviews of Modern Physics}\ }\textbf {\bibinfo {volume}
  {81}},\ \bibinfo {pages} {163} (\bibinfo {year} {2009})}\BibitemShut
  {NoStop}%
\bibitem [{\citenamefont {Krausz}\ and\ \citenamefont
  {Stockman}(2014)}]{Stockman2014}%
  \BibitemOpen
  \bibfield  {author} {\bibinfo {author} {\bibfnamefont {F.}~\bibnamefont
  {Krausz}}\ and\ \bibinfo {author} {\bibfnamefont {M.~I.}\ \bibnamefont
  {Stockman}},\ }\href {\doibase 10.1038/nphoton.2014.28} {\bibfield  {journal}
  {\bibinfo  {journal} {Nature Photonics}\ }\textbf {\bibinfo {volume} {8}},\
  \bibinfo {pages} {205} (\bibinfo {year} {2014})}\BibitemShut {NoStop}%
\bibitem [{\citenamefont {Chang}\ \emph {et~al.}(2016)\citenamefont {Chang},
  \citenamefont {Corkum},\ and\ \citenamefont {Leone}}]{Chang2016}%
  \BibitemOpen
  \bibfield  {author} {\bibinfo {author} {\bibfnamefont {Z.}~\bibnamefont
  {Chang}}, \bibinfo {author} {\bibfnamefont {P.~B.}\ \bibnamefont {Corkum}}, \
  and\ \bibinfo {author} {\bibfnamefont {S.~R.}\ \bibnamefont {Leone}},\ }\href
  {\doibase 10.1364/josab.33.001081} {\bibfield  {journal} {\bibinfo  {journal}
  {Journal of the Optical Society of America B}\ }\textbf {\bibinfo {volume}
  {33}},\ \bibinfo {pages} {1081} (\bibinfo {year} {2016})}\BibitemShut
  {NoStop}%
\bibitem [{\citenamefont {Nisoli}\ \emph {et~al.}(2017)\citenamefont {Nisoli},
  \citenamefont {Decleva}, \citenamefont {Calegari}, \citenamefont {Palacios},\
  and\ \citenamefont {Mart{\'{i}}n}}]{Nisoli2017}%
  \BibitemOpen
  \bibfield  {author} {\bibinfo {author} {\bibfnamefont {M.}~\bibnamefont
  {Nisoli}}, \bibinfo {author} {\bibfnamefont {P.}~\bibnamefont {Decleva}},
  \bibinfo {author} {\bibfnamefont {F.}~\bibnamefont {Calegari}}, \bibinfo
  {author} {\bibfnamefont {A.}~\bibnamefont {Palacios}}, \ and\ \bibinfo
  {author} {\bibfnamefont {F.}~\bibnamefont {Mart{\'{i}}n}},\ }\href {\doibase
  10.1021/acs.chemrev.6b00453} {\enquote {\bibinfo {title} {{Attosecond
  Electron Dynamics in Molecules}},}\ } (\bibinfo {year} {2017})\BibitemShut
  {NoStop}%
\bibitem [{\citenamefont {Lin}\ \emph {et~al.}(2018)\citenamefont {Lin},
  \citenamefont {Le}, \citenamefont {Jin},\ and\ \citenamefont
  {Wei}}]{Lin2018}%
  \BibitemOpen
  \bibfield  {author} {\bibinfo {author} {\bibfnamefont {C.~D.}\ \bibnamefont
  {Lin}}, \bibinfo {author} {\bibfnamefont {A.-T.}\ \bibnamefont {Le}},
  \bibinfo {author} {\bibfnamefont {C.}~\bibnamefont {Jin}}, \ and\ \bibinfo
  {author} {\bibfnamefont {H.}~\bibnamefont {Wei}},\ }\href {\doibase
  10.1017/9781108181839} {\emph {\bibinfo {title} {Attosecond and Strong-Field
  Physics}}}\ (\bibinfo  {publisher} {Cambridge University Press},\ \bibinfo
  {year} {2018})\BibitemShut {NoStop}%
\bibitem [{\citenamefont {Zeng}\ \emph {et~al.}(2007)\citenamefont {Zeng},
  \citenamefont {Cheng}, \citenamefont {Song}, \citenamefont {Li},\ and\
  \citenamefont {Xu}}]{Zeng2007}%
  \BibitemOpen
  \bibfield  {author} {\bibinfo {author} {\bibfnamefont {Z.}~\bibnamefont
  {Zeng}}, \bibinfo {author} {\bibfnamefont {Y.}~\bibnamefont {Cheng}},
  \bibinfo {author} {\bibfnamefont {X.}~\bibnamefont {Song}}, \bibinfo {author}
  {\bibfnamefont {R.}~\bibnamefont {Li}}, \ and\ \bibinfo {author}
  {\bibfnamefont {Z.}~\bibnamefont {Xu}},\ }\href {\doibase
  10.1103/PhysRevLett.98.203901} {\bibfield  {journal} {\bibinfo  {journal}
  {Physical Review Letters}\ }\textbf {\bibinfo {volume} {98}},\ \bibinfo
  {pages} {203901} (\bibinfo {year} {2007})}\BibitemShut {NoStop}%
\bibitem [{\citenamefont {Popmintchev}\ \emph {et~al.}(2010)\citenamefont
  {Popmintchev}, \citenamefont {Chen}, \citenamefont {Arpin}, \citenamefont
  {Murnane},\ and\ \citenamefont {Kapteyn}}]{Popmintchev2010}%
  \BibitemOpen
  \bibfield  {author} {\bibinfo {author} {\bibfnamefont {T.}~\bibnamefont
  {Popmintchev}}, \bibinfo {author} {\bibfnamefont {M.~C.}\ \bibnamefont
  {Chen}}, \bibinfo {author} {\bibfnamefont {P.}~\bibnamefont {Arpin}},
  \bibinfo {author} {\bibfnamefont {M.~M.}\ \bibnamefont {Murnane}}, \ and\
  \bibinfo {author} {\bibfnamefont {H.~C.}\ \bibnamefont {Kapteyn}},\ }\href
  {\doibase 10.1038/nphoton.2010.256} {\enquote {\bibinfo {title} {{The
  attosecond nonlinear optics of bright coherent X-ray generation}},}\ }
  (\bibinfo {year} {2010})\BibitemShut {NoStop}%
\bibitem [{\citenamefont {Popmintchev}\ \emph {et~al.}(2012)\citenamefont
  {Popmintchev}, \citenamefont {Chen}, \citenamefont {Popmintchev},
  \citenamefont {Arpin}, \citenamefont {Brown}, \citenamefont
  {Ali{\v{s}}auskas}, \citenamefont {Andriukaitis}, \citenamefont
  {Bal{\v{c}}iunas}, \citenamefont {M{\"{u}}cke}, \citenamefont {Pugzlys},
  \citenamefont {Baltu{\v{s}}ka}, \citenamefont {Shim}, \citenamefont
  {Schrauth}, \citenamefont {Gaeta}, \citenamefont
  {Hern{\'{a}}ndez-Garc{\'{i}}a}, \citenamefont {Plaja}, \citenamefont
  {Becker}, \citenamefont {Jaron-Becker}, \citenamefont {Murnane},\ and\
  \citenamefont {Kapteyn}}]{Popmintchev2012}%
  \BibitemOpen
  \bibfield  {author} {\bibinfo {author} {\bibfnamefont {T.}~\bibnamefont
  {Popmintchev}}, \bibinfo {author} {\bibfnamefont {M.~C.}\ \bibnamefont
  {Chen}}, \bibinfo {author} {\bibfnamefont {D.}~\bibnamefont {Popmintchev}},
  \bibinfo {author} {\bibfnamefont {P.}~\bibnamefont {Arpin}}, \bibinfo
  {author} {\bibfnamefont {S.}~\bibnamefont {Brown}}, \bibinfo {author}
  {\bibfnamefont {S.}~\bibnamefont {Ali{\v{s}}auskas}}, \bibinfo {author}
  {\bibfnamefont {G.}~\bibnamefont {Andriukaitis}}, \bibinfo {author}
  {\bibfnamefont {T.}~\bibnamefont {Bal{\v{c}}iunas}}, \bibinfo {author}
  {\bibfnamefont {O.~D.}\ \bibnamefont {M{\"{u}}cke}}, \bibinfo {author}
  {\bibfnamefont {A.}~\bibnamefont {Pugzlys}}, \bibinfo {author} {\bibfnamefont
  {A.}~\bibnamefont {Baltu{\v{s}}ka}}, \bibinfo {author} {\bibfnamefont
  {B.}~\bibnamefont {Shim}}, \bibinfo {author} {\bibfnamefont {S.~E.}\
  \bibnamefont {Schrauth}}, \bibinfo {author} {\bibfnamefont {A.}~\bibnamefont
  {Gaeta}}, \bibinfo {author} {\bibfnamefont {C.}~\bibnamefont
  {Hern{\'{a}}ndez-Garc{\'{i}}a}}, \bibinfo {author} {\bibfnamefont
  {L.}~\bibnamefont {Plaja}}, \bibinfo {author} {\bibfnamefont
  {A.}~\bibnamefont {Becker}}, \bibinfo {author} {\bibfnamefont
  {A.}~\bibnamefont {Jaron-Becker}}, \bibinfo {author} {\bibfnamefont {M.~M.}\
  \bibnamefont {Murnane}}, \ and\ \bibinfo {author} {\bibfnamefont {H.~C.}\
  \bibnamefont {Kapteyn}},\ }\href {\doibase 10.1126/science.1218497}
  {\bibfield  {journal} {\bibinfo  {journal} {Science}\ }\textbf {\bibinfo
  {volume} {336}},\ \bibinfo {pages} {1287} (\bibinfo {year}
  {2012})}\BibitemShut {NoStop}%
\bibitem [{\citenamefont {Seres}\ \emph {et~al.}(2005)\citenamefont {Seres},
  \citenamefont {Seres}, \citenamefont {Verhoef}, \citenamefont {Tempea},
  \citenamefont {Streli}, \citenamefont {Wobrauschek}, \citenamefont
  {Yakovlev}, \citenamefont {Scrinzi}, \citenamefont {Spielmann},\ and\
  \citenamefont {Krausz}}]{Seres2005}%
  \BibitemOpen
  \bibfield  {author} {\bibinfo {author} {\bibfnamefont {J.}~\bibnamefont
  {Seres}}, \bibinfo {author} {\bibfnamefont {E.}~\bibnamefont {Seres}},
  \bibinfo {author} {\bibfnamefont {A.~J.}\ \bibnamefont {Verhoef}}, \bibinfo
  {author} {\bibfnamefont {G.}~\bibnamefont {Tempea}}, \bibinfo {author}
  {\bibfnamefont {C.}~\bibnamefont {Streli}}, \bibinfo {author} {\bibfnamefont
  {P.}~\bibnamefont {Wobrauschek}}, \bibinfo {author} {\bibfnamefont
  {V.}~\bibnamefont {Yakovlev}}, \bibinfo {author} {\bibfnamefont
  {A.}~\bibnamefont {Scrinzi}}, \bibinfo {author} {\bibfnamefont
  {C.}~\bibnamefont {Spielmann}}, \ and\ \bibinfo {author} {\bibfnamefont
  {F.}~\bibnamefont {Krausz}},\ }\href {\doibase 10.1038/433596a} {\bibfield
  {journal} {\bibinfo  {journal} {Nature}\ }\textbf {\bibinfo {volume} {433}},\
  \bibinfo {pages} {596} (\bibinfo {year} {2005})}\BibitemShut {NoStop}%
\bibitem [{\citenamefont {Ishikawa}\ \emph {et~al.}(2012)\citenamefont
  {Ishikawa} \emph {et~al.}}]{Ishikawa2012}%
  \BibitemOpen
  \bibfield  {author} {\bibinfo {author} {\bibfnamefont {T.}~\bibnamefont
  {Ishikawa}} \emph {et~al.},\ }\href {\doibase 10.1038/nphoton.2012.141}
  {\bibfield  {journal} {\bibinfo  {journal} {Nature Photonics}\ }\textbf
  {\bibinfo {volume} {6}},\ \bibinfo {pages} {540} (\bibinfo {year}
  {2012})}\BibitemShut {NoStop}%
\bibitem [{\citenamefont {Ackermann}\ \emph {et~al.}(2007)\citenamefont
  {Ackermann} \emph {et~al.}}]{Ackermann2007}%
  \BibitemOpen
  \bibfield  {author} {\bibinfo {author} {\bibfnamefont {W.}~\bibnamefont
  {Ackermann}} \emph {et~al.},\ }\href {\doibase 10.1038/nphoton.2007.76}
  {\bibfield  {journal} {\bibinfo  {journal} {Nature Photonics}\ }\textbf
  {\bibinfo {volume} {1}},\ \bibinfo {pages} {336} (\bibinfo {year}
  {2007})}\BibitemShut {NoStop}%
\bibitem [{\citenamefont {Popmintchev}\ \emph {et~al.}(2009)\citenamefont
  {Popmintchev}, \citenamefont {Chen}, \citenamefont {Bahabad}, \citenamefont
  {Gerrity}, \citenamefont {Sidorenko}, \citenamefont {Cohen}, \citenamefont
  {Christov}, \citenamefont {Murnane},\ and\ \citenamefont
  {Kapteyn}}]{Popmintchev2009}%
  \BibitemOpen
  \bibfield  {author} {\bibinfo {author} {\bibfnamefont {T.}~\bibnamefont
  {Popmintchev}}, \bibinfo {author} {\bibfnamefont {M.-C.}\ \bibnamefont
  {Chen}}, \bibinfo {author} {\bibfnamefont {A.}~\bibnamefont {Bahabad}},
  \bibinfo {author} {\bibfnamefont {M.}~\bibnamefont {Gerrity}}, \bibinfo
  {author} {\bibfnamefont {P.}~\bibnamefont {Sidorenko}}, \bibinfo {author}
  {\bibfnamefont {O.}~\bibnamefont {Cohen}}, \bibinfo {author} {\bibfnamefont
  {I.~P.}\ \bibnamefont {Christov}}, \bibinfo {author} {\bibfnamefont {M.~M.}\
  \bibnamefont {Murnane}}, \ and\ \bibinfo {author} {\bibfnamefont {H.~C.}\
  \bibnamefont {Kapteyn}},\ }\href {\doibase 10.1073/pnas.0903748106}
  {\bibfield  {journal} {\bibinfo  {journal} {Proceedings of the National
  Academy of Sciences}\ }\textbf {\bibinfo {volume} {106}},\ \bibinfo {pages}
  {10516} (\bibinfo {year} {2009})}\BibitemShut {NoStop}%
\bibitem [{\citenamefont {Sansone}\ \emph {et~al.}(2006)\citenamefont
  {Sansone}, \citenamefont {Benedetti}, \citenamefont {Calegari}, \citenamefont
  {Vozzi}, \citenamefont {Avaldi}, \citenamefont {Flammini}, \citenamefont
  {Poletto}, \citenamefont {Villoresi}, \citenamefont {Altucci}, \citenamefont
  {Velotta}, \citenamefont {Stagira}, \citenamefont {{De Silvestri}},\ and\
  \citenamefont {Nisoli}}]{Sansone2006}%
  \BibitemOpen
  \bibfield  {author} {\bibinfo {author} {\bibfnamefont {G.}~\bibnamefont
  {Sansone}}, \bibinfo {author} {\bibfnamefont {E.}~\bibnamefont {Benedetti}},
  \bibinfo {author} {\bibfnamefont {F.}~\bibnamefont {Calegari}}, \bibinfo
  {author} {\bibfnamefont {C.}~\bibnamefont {Vozzi}}, \bibinfo {author}
  {\bibfnamefont {L.}~\bibnamefont {Avaldi}}, \bibinfo {author} {\bibfnamefont
  {R.}~\bibnamefont {Flammini}}, \bibinfo {author} {\bibfnamefont
  {L.}~\bibnamefont {Poletto}}, \bibinfo {author} {\bibfnamefont
  {P.}~\bibnamefont {Villoresi}}, \bibinfo {author} {\bibfnamefont
  {C.}~\bibnamefont {Altucci}}, \bibinfo {author} {\bibfnamefont
  {R.}~\bibnamefont {Velotta}}, \bibinfo {author} {\bibfnamefont
  {S.}~\bibnamefont {Stagira}}, \bibinfo {author} {\bibfnamefont
  {S.}~\bibnamefont {{De Silvestri}}}, \ and\ \bibinfo {author} {\bibfnamefont
  {M.}~\bibnamefont {Nisoli}},\ }\href {\doibase 10.1126/science.1132838}
  {\bibfield  {journal} {\bibinfo  {journal} {Science}\ }\textbf {\bibinfo
  {volume} {314}},\ \bibinfo {pages} {443} (\bibinfo {year}
  {2006})}\BibitemShut {NoStop}%
\bibitem [{\citenamefont {Feng}\ \emph {et~al.}(2009)\citenamefont {Feng},
  \citenamefont {Gilbertson}, \citenamefont {Mashiko}, \citenamefont {Wang},
  \citenamefont {Khan}, \citenamefont {Chini}, \citenamefont {Wu},
  \citenamefont {Zhao},\ and\ \citenamefont {Chang}}]{Feng2009}%
  \BibitemOpen
  \bibfield  {author} {\bibinfo {author} {\bibfnamefont {X.}~\bibnamefont
  {Feng}}, \bibinfo {author} {\bibfnamefont {S.}~\bibnamefont {Gilbertson}},
  \bibinfo {author} {\bibfnamefont {H.}~\bibnamefont {Mashiko}}, \bibinfo
  {author} {\bibfnamefont {H.}~\bibnamefont {Wang}}, \bibinfo {author}
  {\bibfnamefont {S.~D.}\ \bibnamefont {Khan}}, \bibinfo {author}
  {\bibfnamefont {M.}~\bibnamefont {Chini}}, \bibinfo {author} {\bibfnamefont
  {Y.}~\bibnamefont {Wu}}, \bibinfo {author} {\bibfnamefont {K.}~\bibnamefont
  {Zhao}}, \ and\ \bibinfo {author} {\bibfnamefont {Z.}~\bibnamefont {Chang}},\
  }\href {\doibase 10.1103/PhysRevLett.103.183901} {\bibfield  {journal}
  {\bibinfo  {journal} {Physical Review Letters}\ }\textbf {\bibinfo {volume}
  {103}} (\bibinfo {year} {2009}),\ 10.1103/PhysRevLett.103.183901}\BibitemShut
  {NoStop}%
\bibitem [{\citenamefont {Goulielmakis}\ \emph {et~al.}(2008)\citenamefont
  {Goulielmakis}, \citenamefont {Schultze}, \citenamefont {Hofstetter},
  \citenamefont {Yakovlev}, \citenamefont {Gagnon}, \citenamefont {Uiberacker},
  \citenamefont {Aquila}, \citenamefont {Gullikson}, \citenamefont {Attwood},
  \citenamefont {Kienberger}, \citenamefont {Krausz},\ and\ \citenamefont
  {Kleineberg}}]{Goulielmakis2008}%
  \BibitemOpen
  \bibfield  {author} {\bibinfo {author} {\bibfnamefont {E.}~\bibnamefont
  {Goulielmakis}}, \bibinfo {author} {\bibfnamefont {M.}~\bibnamefont
  {Schultze}}, \bibinfo {author} {\bibfnamefont {M.}~\bibnamefont
  {Hofstetter}}, \bibinfo {author} {\bibfnamefont {V.~S.}\ \bibnamefont
  {Yakovlev}}, \bibinfo {author} {\bibfnamefont {J.}~\bibnamefont {Gagnon}},
  \bibinfo {author} {\bibfnamefont {M.}~\bibnamefont {Uiberacker}}, \bibinfo
  {author} {\bibfnamefont {A.~L.}\ \bibnamefont {Aquila}}, \bibinfo {author}
  {\bibfnamefont {E.~M.}\ \bibnamefont {Gullikson}}, \bibinfo {author}
  {\bibfnamefont {D.~T.}\ \bibnamefont {Attwood}}, \bibinfo {author}
  {\bibfnamefont {R.}~\bibnamefont {Kienberger}}, \bibinfo {author}
  {\bibfnamefont {F.}~\bibnamefont {Krausz}}, \ and\ \bibinfo {author}
  {\bibfnamefont {U.}~\bibnamefont {Kleineberg}},\ }\href {\doibase
  10.1126/science.1157846} {\bibfield  {journal} {\bibinfo  {journal}
  {Science}\ }\textbf {\bibinfo {volume} {320}},\ \bibinfo {pages} {1614}
  (\bibinfo {year} {2008})}\BibitemShut {NoStop}%
\bibitem [{\citenamefont {Chini}\ \emph {et~al.}(2014)\citenamefont {Chini},
  \citenamefont {Zhao},\ and\ \citenamefont {Chang}}]{Chini2014}%
  \BibitemOpen
  \bibfield  {author} {\bibinfo {author} {\bibfnamefont {M.}~\bibnamefont
  {Chini}}, \bibinfo {author} {\bibfnamefont {K.}~\bibnamefont {Zhao}}, \ and\
  \bibinfo {author} {\bibfnamefont {Z.}~\bibnamefont {Chang}},\ }\href
  {\doibase 10.1038/nphoton.2013.362} {\bibfield  {journal} {\bibinfo
  {journal} {Nature Photonics}\ }\textbf {\bibinfo {volume} {8}},\ \bibinfo
  {pages} {178} (\bibinfo {year} {2014})}\BibitemShut {NoStop}%
\bibitem [{\citenamefont {Paul}\ \emph {et~al.}(2001)\citenamefont {Paul},
  \citenamefont {Toma}, \citenamefont {Breger}, \citenamefont {Mullot},
  \citenamefont {Aug{\'{e}}}, \citenamefont {Balcou}, \citenamefont {Muller},\
  and\ \citenamefont {Agostini}}]{Paul2001}%
  \BibitemOpen
  \bibfield  {author} {\bibinfo {author} {\bibfnamefont {P.~M.}\ \bibnamefont
  {Paul}}, \bibinfo {author} {\bibfnamefont {E.~S.}\ \bibnamefont {Toma}},
  \bibinfo {author} {\bibfnamefont {P.}~\bibnamefont {Breger}}, \bibinfo
  {author} {\bibfnamefont {G.}~\bibnamefont {Mullot}}, \bibinfo {author}
  {\bibfnamefont {F.}~\bibnamefont {Aug{\'{e}}}}, \bibinfo {author}
  {\bibfnamefont {P.}~\bibnamefont {Balcou}}, \bibinfo {author} {\bibfnamefont
  {H.~G.}\ \bibnamefont {Muller}}, \ and\ \bibinfo {author} {\bibfnamefont
  {P.}~\bibnamefont {Agostini}},\ }\href {\doibase 10.1126/science.1059413}
  {\bibfield  {journal} {\bibinfo  {journal} {Science}\ }\textbf {\bibinfo
  {volume} {292}},\ \bibinfo {pages} {1689} (\bibinfo {year}
  {2001})}\BibitemShut {NoStop}%
\bibitem [{\citenamefont {Winterfeldt}\ \emph {et~al.}(2008)\citenamefont
  {Winterfeldt}, \citenamefont {Spielmann},\ and\ \citenamefont
  {Gerber}}]{Winterfeldt2008}%
  \BibitemOpen
  \bibfield  {author} {\bibinfo {author} {\bibfnamefont {C.}~\bibnamefont
  {Winterfeldt}}, \bibinfo {author} {\bibfnamefont {C.}~\bibnamefont
  {Spielmann}}, \ and\ \bibinfo {author} {\bibfnamefont {G.}~\bibnamefont
  {Gerber}},\ }\href {\doibase 10.1103/RevModPhys.80.117} {\bibfield  {journal}
  {\bibinfo  {journal} {Reviews of Modern Physics}\ }\textbf {\bibinfo {volume}
  {80}},\ \bibinfo {pages} {117} (\bibinfo {year} {2008})}\BibitemShut
  {NoStop}%
\bibitem [{\citenamefont {Corkum}(1993)}]{Corkum1993}%
  \BibitemOpen
  \bibfield  {author} {\bibinfo {author} {\bibfnamefont {P.~B.}\ \bibnamefont
  {Corkum}},\ }\href
  {http://journals.aps.org/prl/abstract/10.1103/PhysRevLett.71.1994} {\bibfield
   {journal} {\bibinfo  {journal} {Physical Review Letters}\ }\textbf {\bibinfo
  {volume} {71}},\ \bibinfo {pages} {1994} (\bibinfo {year}
  {1993})}\BibitemShut {NoStop}%
\bibitem [{\citenamefont {Schafer}\ \emph {et~al.}(1993)\citenamefont
  {Schafer}, \citenamefont {Yang}, \citenamefont {Dimauro},\ and\ \citenamefont
  {Kulander}}]{Schafer1993}%
  \BibitemOpen
  \bibfield  {author} {\bibinfo {author} {\bibfnamefont {K.~J.}\ \bibnamefont
  {Schafer}}, \bibinfo {author} {\bibfnamefont {B.}~\bibnamefont {Yang}},
  \bibinfo {author} {\bibfnamefont {L.~F.}\ \bibnamefont {Dimauro}}, \ and\
  \bibinfo {author} {\bibfnamefont {K.~C.}\ \bibnamefont {Kulander}},\ }\href
  {\doibase 10.1103/PhysRevLett.70.1599} {\bibfield  {journal} {\bibinfo
  {journal} {Physical Review Letters}\ }\textbf {\bibinfo {volume} {70}},\
  \bibinfo {pages} {1599} (\bibinfo {year} {1993})}\BibitemShut {NoStop}%
\bibitem [{\citenamefont {Krause}\ \emph
  {et~al.}(1992{\natexlab{a}})\citenamefont {Krause}, \citenamefont {Schafer},\
  and\ \citenamefont {Kulander}}]{Krause1992a}%
  \BibitemOpen
  \bibfield  {author} {\bibinfo {author} {\bibfnamefont {J.~L.}\ \bibnamefont
  {Krause}}, \bibinfo {author} {\bibfnamefont {K.~J.}\ \bibnamefont {Schafer}},
  \ and\ \bibinfo {author} {\bibfnamefont {K.~C.}\ \bibnamefont {Kulander}},\
  }\href {\doibase 10.1103/PhysRevLett.68.3535} {\bibfield  {journal} {\bibinfo
   {journal} {Physical Review Letters}\ }\textbf {\bibinfo {volume} {68}},\
  \bibinfo {pages} {3535} (\bibinfo {year} {1992}{\natexlab{a}})}\BibitemShut
  {NoStop}%
\bibitem [{\citenamefont {Lewenstein}\ \emph {et~al.}(1994)\citenamefont
  {Lewenstein}, \citenamefont {Balcou}, \citenamefont {Ivanov}, \citenamefont
  {L'Huillier},\ and\ \citenamefont {Corkum}}]{Lewenstein1994}%
  \BibitemOpen
  \bibfield  {author} {\bibinfo {author} {\bibfnamefont {M.}~\bibnamefont
  {Lewenstein}}, \bibinfo {author} {\bibfnamefont {P.}~\bibnamefont {Balcou}},
  \bibinfo {author} {\bibfnamefont {M.~Y.}\ \bibnamefont {Ivanov}}, \bibinfo
  {author} {\bibfnamefont {A.}~\bibnamefont {L'Huillier}}, \ and\ \bibinfo
  {author} {\bibfnamefont {P.~B.}\ \bibnamefont {Corkum}},\ }\href {\doibase
  10.1103/PhysRevA.49.2117} {\bibfield  {journal} {\bibinfo  {journal}
  {Physical Review A}\ }\textbf {\bibinfo {volume} {49}},\ \bibinfo {pages}
  {2117} (\bibinfo {year} {1994})}\BibitemShut {NoStop}%
\bibitem [{\citenamefont {Baier}\ \emph {et~al.}(2006)\citenamefont {Baier},
  \citenamefont {Ruiz}, \citenamefont {Plaja},\ and\ \citenamefont
  {Becker}}]{Baier2006}%
  \BibitemOpen
  \bibfield  {author} {\bibinfo {author} {\bibfnamefont {S.}~\bibnamefont
  {Baier}}, \bibinfo {author} {\bibfnamefont {C.}~\bibnamefont {Ruiz}},
  \bibinfo {author} {\bibfnamefont {L.}~\bibnamefont {Plaja}}, \ and\ \bibinfo
  {author} {\bibfnamefont {A.}~\bibnamefont {Becker}},\ }\href {\doibase
  10.1103/PhysRevA.74.033405} {\bibfield  {journal} {\bibinfo  {journal}
  {Physical Review A - Atomic, Molecular, and Optical Physics}\ }\textbf
  {\bibinfo {volume} {74}} (\bibinfo {year} {2006}),\
  10.1103/PhysRevA.74.033405}\BibitemShut {NoStop}%
\bibitem [{\citenamefont {Baier}\ \emph {et~al.}(2007)\citenamefont {Baier},
  \citenamefont {Ruiz}, \citenamefont {Plaja},\ and\ \citenamefont
  {Becker}}]{Baier2007}%
  \BibitemOpen
  \bibfield  {author} {\bibinfo {author} {\bibfnamefont {S.}~\bibnamefont
  {Baier}}, \bibinfo {author} {\bibfnamefont {C.}~\bibnamefont {Ruiz}},
  \bibinfo {author} {\bibfnamefont {L.}~\bibnamefont {Plaja}}, \ and\ \bibinfo
  {author} {\bibfnamefont {A.}~\bibnamefont {Becker}},\ }\href {\doibase
  10.1134/s1054660x07040111} {\bibfield  {journal} {\bibinfo  {journal} {Laser
  Physics}\ }\textbf {\bibinfo {volume} {17}},\ \bibinfo {pages} {358}
  (\bibinfo {year} {2007})}\BibitemShut {NoStop}%
\bibitem [{\citenamefont {Chelkowski}\ \emph {et~al.}(2012)\citenamefont
  {Chelkowski}, \citenamefont {Bredtmann},\ and\ \citenamefont
  {Bandrauk}}]{Chelkowski2012}%
  \BibitemOpen
  \bibfield  {author} {\bibinfo {author} {\bibfnamefont {S.}~\bibnamefont
  {Chelkowski}}, \bibinfo {author} {\bibfnamefont {T.}~\bibnamefont
  {Bredtmann}}, \ and\ \bibinfo {author} {\bibfnamefont {A.~D.}\ \bibnamefont
  {Bandrauk}},\ }\href {\doibase 10.1103/PhysRevA.85.033404} {\bibfield
  {journal} {\bibinfo  {journal} {Physical Review A - Atomic, Molecular, and
  Optical Physics}\ }\textbf {\bibinfo {volume} {85}},\ \bibinfo {pages}
  {33404} (\bibinfo {year} {2012})}\BibitemShut {NoStop}%
\bibitem [{\citenamefont {Morales}\ \emph {et~al.}(2014)\citenamefont
  {Morales}, \citenamefont {Rivi{\`{e}}re}, \citenamefont {Richter},
  \citenamefont {Gubaydullin}, \citenamefont {Ivanov}, \citenamefont
  {Smirnova},\ and\ \citenamefont {Mart{\'{i}}n}}]{Morales2014}%
  \BibitemOpen
  \bibfield  {author} {\bibinfo {author} {\bibfnamefont {F.}~\bibnamefont
  {Morales}}, \bibinfo {author} {\bibfnamefont {P.}~\bibnamefont
  {Rivi{\`{e}}re}}, \bibinfo {author} {\bibfnamefont {M.}~\bibnamefont
  {Richter}}, \bibinfo {author} {\bibfnamefont {A.}~\bibnamefont
  {Gubaydullin}}, \bibinfo {author} {\bibfnamefont {M.}~\bibnamefont {Ivanov}},
  \bibinfo {author} {\bibfnamefont {O.}~\bibnamefont {Smirnova}}, \ and\
  \bibinfo {author} {\bibfnamefont {F.}~\bibnamefont {Mart{\'{i}}n}},\ }\href
  {\doibase 10.1088/0953-4075/47/20/204015} {\bibfield  {journal} {\bibinfo
  {journal} {Journal of Physics B: Atomic, Molecular and Optical Physics}\
  }\textbf {\bibinfo {volume} {47}} (\bibinfo {year} {2014}),\
  10.1088/0953-4075/47/20/204015}\BibitemShut {NoStop}%
\bibitem [{\citenamefont {Su{\'{a}}rez}\ \emph {et~al.}(2017)\citenamefont
  {Su{\'{a}}rez}, \citenamefont {Chac{\'{o}}n}, \citenamefont
  {P{\'{e}}rez-Hern{\'{a}}ndez}, \citenamefont {Biegert}, \citenamefont
  {Lewenstein},\ and\ \citenamefont {Ciappina}}]{Suarez2017}%
  \BibitemOpen
  \bibfield  {author} {\bibinfo {author} {\bibfnamefont {N.}~\bibnamefont
  {Su{\'{a}}rez}}, \bibinfo {author} {\bibfnamefont {A.}~\bibnamefont
  {Chac{\'{o}}n}}, \bibinfo {author} {\bibfnamefont {J.~A.}\ \bibnamefont
  {P{\'{e}}rez-Hern{\'{a}}ndez}}, \bibinfo {author} {\bibfnamefont
  {J.}~\bibnamefont {Biegert}}, \bibinfo {author} {\bibfnamefont
  {M.}~\bibnamefont {Lewenstein}}, \ and\ \bibinfo {author} {\bibfnamefont
  {M.~F.}\ \bibnamefont {Ciappina}},\ }\href {\doibase
  10.1103/PhysRevA.95.033415} {\bibfield  {journal} {\bibinfo  {journal}
  {Physical Review A}\ }\textbf {\bibinfo {volume} {95}},\ \bibinfo {pages}
  {33415} (\bibinfo {year} {2017})}\BibitemShut {NoStop}%
\bibitem [{\citenamefont {Lewenstein}\ \emph {et~al.}(1995)\citenamefont
  {Lewenstein}, \citenamefont {Kulander}, \citenamefont {Schafer},
  \citenamefont {Bucksbaum}, \citenamefont {{Fizyki Teoretycznej}},\ and\
  \citenamefont {{Akademia Nauk}}}]{Lewenstein1995}%
  \BibitemOpen
  \bibfield  {author} {\bibinfo {author} {\bibfnamefont {M.}~\bibnamefont
  {Lewenstein}}, \bibinfo {author} {\bibfnamefont {K.~C.}\ \bibnamefont
  {Kulander}}, \bibinfo {author} {\bibfnamefont {K.~J.}\ \bibnamefont
  {Schafer}}, \bibinfo {author} {\bibfnamefont {P.~H.}\ \bibnamefont
  {Bucksbaum}}, \bibinfo {author} {\bibfnamefont {C.}~\bibnamefont {{Fizyki
  Teoretycznej}}}, \ and\ \bibinfo {author} {\bibfnamefont {P.}~\bibnamefont
  {{Akademia Nauk}}},\ }\href
  {https://journals.aps.org/pra/pdf/10.1103/PhysRevA.51.1495} {\emph {\bibinfo
  {title} {PHYSICAL REVIEW A}}},\ \bibinfo {type} {Tech. Rep.}\ \bibinfo
  {number} {2}\ (\bibinfo {year} {1995})\BibitemShut {NoStop}%
\bibitem [{\citenamefont {Lu}\ \emph {et~al.}(2008)\citenamefont {Lu},
  \citenamefont {Zhang},\ and\ \citenamefont {Han}}]{Lu2008}%
  \BibitemOpen
  \bibfield  {author} {\bibinfo {author} {\bibfnamefont {R.~F.}\ \bibnamefont
  {Lu}}, \bibinfo {author} {\bibfnamefont {P.~Y.}\ \bibnamefont {Zhang}}, \
  and\ \bibinfo {author} {\bibfnamefont {K.~L.}\ \bibnamefont {Han}},\ }\href
  {\doibase 10.1103/PhysRevE.77.066701} {\bibfield  {journal} {\bibinfo
  {journal} {Physical Review E - Statistical, Nonlinear, and Soft Matter
  Physics}\ }\textbf {\bibinfo {volume} {77}} (\bibinfo {year} {2008}),\
  10.1103/PhysRevE.77.066701}\BibitemShut {NoStop}%
\bibitem [{\citenamefont {Chen}\ \emph {et~al.}(2006)\citenamefont {Chen},
  \citenamefont {Morishita}, \citenamefont {Le}, \citenamefont {Wickenhauser},
  \citenamefont {Tong},\ and\ \citenamefont {Lin}}]{Chen2006}%
  \BibitemOpen
  \bibfield  {author} {\bibinfo {author} {\bibfnamefont {Z.}~\bibnamefont
  {Chen}}, \bibinfo {author} {\bibfnamefont {T.}~\bibnamefont {Morishita}},
  \bibinfo {author} {\bibfnamefont {A.~T.}\ \bibnamefont {Le}}, \bibinfo
  {author} {\bibfnamefont {M.}~\bibnamefont {Wickenhauser}}, \bibinfo {author}
  {\bibfnamefont {X.~M.}\ \bibnamefont {Tong}}, \ and\ \bibinfo {author}
  {\bibfnamefont {C.~D.}\ \bibnamefont {Lin}},\ }\href {\doibase
  10.1103/PhysRevA.74.053405} {\bibfield  {journal} {\bibinfo  {journal}
  {Physical Review A - Atomic, Molecular, and Optical Physics}\ }\textbf
  {\bibinfo {volume} {74}} (\bibinfo {year} {2006}),\
  10.1103/PhysRevA.74.053405}\BibitemShut {NoStop}%
\bibitem [{\citenamefont {Morishita}\ \emph {et~al.}(2007)\citenamefont
  {Morishita}, \citenamefont {Chen}, \citenamefont {Watanabe},\ and\
  \citenamefont {Lin}}]{Morishita2007}%
  \BibitemOpen
  \bibfield  {author} {\bibinfo {author} {\bibfnamefont {T.}~\bibnamefont
  {Morishita}}, \bibinfo {author} {\bibfnamefont {Z.}~\bibnamefont {Chen}},
  \bibinfo {author} {\bibfnamefont {S.}~\bibnamefont {Watanabe}}, \ and\
  \bibinfo {author} {\bibfnamefont {C.~D.}\ \bibnamefont {Lin}},\ }\href
  {\doibase 10.1103/PhysRevA.75.023407} {\bibfield  {journal} {\bibinfo
  {journal} {Physical Review A - Atomic, Molecular, and Optical Physics}\
  }\textbf {\bibinfo {volume} {75}} (\bibinfo {year} {2007}),\
  10.1103/PhysRevA.75.023407}\BibitemShut {NoStop}%
\bibitem [{\citenamefont {Bauer}\ and\ \citenamefont
  {Koval}(2006)}]{Bauer2006}%
  \BibitemOpen
  \bibfield  {author} {\bibinfo {author} {\bibfnamefont {D.}~\bibnamefont
  {Bauer}}\ and\ \bibinfo {author} {\bibfnamefont {P.}~\bibnamefont {Koval}},\
  }\href {\doibase 10.1016/j.cpc.2005.11.001} {\bibfield  {journal} {\bibinfo
  {journal} {Computer Physics Communications}\ }\textbf {\bibinfo {volume}
  {174}},\ \bibinfo {pages} {396} (\bibinfo {year} {2006})}\BibitemShut
  {NoStop}%
\bibitem [{\citenamefont {Zhou}\ and\ \citenamefont
  {Chu}(2011{\natexlab{a}})}]{pseudospectral_Zhou2011}%
  \BibitemOpen
  \bibfield  {author} {\bibinfo {author} {\bibfnamefont {Z.}~\bibnamefont
  {Zhou}}\ and\ \bibinfo {author} {\bibfnamefont {S.~I.}\ \bibnamefont {Chu}},\
  }\href {\doibase 10.1103/PhysRevA.83.013405} {\bibfield  {journal} {\bibinfo
  {journal} {Physical Review A - Atomic, Molecular, and Optical Physics}\
  }\textbf {\bibinfo {volume} {83}},\ \bibinfo {pages} {13405} (\bibinfo {year}
  {2011}{\natexlab{a}})}\BibitemShut {NoStop}%
\bibitem [{\citenamefont {Sanz-Vicario}\ \emph {et~al.}(2006)\citenamefont
  {Sanz-Vicario}, \citenamefont {Bachau},\ and\ \citenamefont
  {Mart{\'{i}}n}}]{Sanz-Vicario2006}%
  \BibitemOpen
  \bibfield  {author} {\bibinfo {author} {\bibfnamefont {J.~L.}\ \bibnamefont
  {Sanz-Vicario}}, \bibinfo {author} {\bibfnamefont {H.}~\bibnamefont
  {Bachau}}, \ and\ \bibinfo {author} {\bibfnamefont {F.}~\bibnamefont
  {Mart{\'{i}}n}},\ }\href {\doibase 10.1103/PhysRevA.73.033410} {\bibfield
  {journal} {\bibinfo  {journal} {Physical Review A - Atomic, Molecular, and
  Optical Physics}\ }\textbf {\bibinfo {volume} {73}} (\bibinfo {year}
  {2006}),\ 10.1103/PhysRevA.73.033410}\BibitemShut {NoStop}%
\bibitem [{\citenamefont {Zatsarinny}(2006)}]{Zatsarinny2006}%
  \BibitemOpen
  \bibfield  {author} {\bibinfo {author} {\bibfnamefont {O.}~\bibnamefont
  {Zatsarinny}},\ }\href {\doibase 10.1016/j.cpc.2005.10.006} {\bibfield
  {journal} {\bibinfo  {journal} {Computer Physics Communications}\ }\textbf
  {\bibinfo {volume} {174}},\ \bibinfo {pages} {273} (\bibinfo {year}
  {2006})}\BibitemShut {NoStop}%
\bibitem [{\citenamefont {Awasthi}\ \emph {et~al.}(2005)\citenamefont
  {Awasthi}, \citenamefont {Vanne},\ and\ \citenamefont {Saenz}}]{Awasthi2005}%
  \BibitemOpen
  \bibfield  {author} {\bibinfo {author} {\bibfnamefont {M.}~\bibnamefont
  {Awasthi}}, \bibinfo {author} {\bibfnamefont {Y.~V.}\ \bibnamefont {Vanne}},
  \ and\ \bibinfo {author} {\bibfnamefont {A.}~\bibnamefont {Saenz}},\ }\href
  {\doibase 10.1088/0953-4075/38/22/005} {\bibfield  {journal} {\bibinfo
  {journal} {Journal of Physics B: Atomic, Molecular and Optical Physics}\
  }\textbf {\bibinfo {volume} {38}},\ \bibinfo {pages} {3973} (\bibinfo {year}
  {2005})}\BibitemShut {NoStop}%
\bibitem [{\citenamefont {Meyer}(2012)}]{Meyer2012}%
  \BibitemOpen
  \bibfield  {author} {\bibinfo {author} {\bibfnamefont {H.~D.}\ \bibnamefont
  {Meyer}},\ }\href {\doibase 10.1002/wcms.87} {\enquote {\bibinfo {title}
  {{Studying molecular quantum dynamics with the multiconfiguration
  time-dependent Hartree method}},}\ } (\bibinfo {year} {2012})\BibitemShut
  {NoStop}%
\bibitem [{\citenamefont {Caillat}\ \emph {et~al.}(2005)\citenamefont
  {Caillat}, \citenamefont {Zanghellini}, \citenamefont {Kitzler},
  \citenamefont {Koch}, \citenamefont {Kreuzer},\ and\ \citenamefont
  {Scrinzi}}]{Caillat_Scrinzi2005}%
  \BibitemOpen
  \bibfield  {author} {\bibinfo {author} {\bibfnamefont {J.}~\bibnamefont
  {Caillat}}, \bibinfo {author} {\bibfnamefont {J.}~\bibnamefont
  {Zanghellini}}, \bibinfo {author} {\bibfnamefont {M.}~\bibnamefont
  {Kitzler}}, \bibinfo {author} {\bibfnamefont {O.}~\bibnamefont {Koch}},
  \bibinfo {author} {\bibfnamefont {W.}~\bibnamefont {Kreuzer}}, \ and\
  \bibinfo {author} {\bibfnamefont {A.}~\bibnamefont {Scrinzi}},\ }\href
  {\doibase 10.1103/PhysRevA.71.012712} {\bibfield  {journal} {\bibinfo
  {journal} {Physical Review A - Atomic, Molecular, and Optical Physics}\
  }\textbf {\bibinfo {volume} {71}} (\bibinfo {year} {2005}),\
  10.1103/PhysRevA.71.012712}\BibitemShut {NoStop}%
\bibitem [{\citenamefont {Casida}\ and\ \citenamefont
  {Huix-Rotllant}(2012)}]{Casida2012}%
  \BibitemOpen
  \bibfield  {author} {\bibinfo {author} {\bibfnamefont {M.}~\bibnamefont
  {Casida}}\ and\ \bibinfo {author} {\bibfnamefont {M.}~\bibnamefont
  {Huix-Rotllant}},\ }\href {\doibase 10.1146/annurev-physchem-032511-143803}
  {\bibfield  {journal} {\bibinfo  {journal} {Annual Review of Physical
  Chemistry}\ }\textbf {\bibinfo {volume} {63}},\ \bibinfo {pages} {287}
  (\bibinfo {year} {2012})}\BibitemShut {NoStop}%
\bibitem [{\citenamefont {Castro}\ \emph {et~al.}(2012)\citenamefont {Castro},
  \citenamefont {Gross}, \citenamefont {Ruggenthaler}, \citenamefont
  {Leeuwen},\ and\ \citenamefont {Dobson}}]{Castro2012}%
  \BibitemOpen
  \bibfield  {author} {\bibinfo {author} {\bibfnamefont {A.}~\bibnamefont
  {Castro}}, \bibinfo {author} {\bibfnamefont {E.~K.~U.}\ \bibnamefont
  {Gross}}, \bibinfo {author} {\bibfnamefont {M.}~\bibnamefont {Ruggenthaler}},
  \bibinfo {author} {\bibfnamefont {R.~V.}\ \bibnamefont {Leeuwen}}, \ and\
  \bibinfo {author} {\bibfnamefont {J.~F.}\ \bibnamefont {Dobson}},\ }\href
  {\doibase 10.1007/978-3-642-23518-4} {\bibfield  {journal} {\bibinfo
  {journal} {Lecture Notes in Physics}\ }\textbf {\bibinfo {volume} {837}},\
  \bibinfo {pages} {417} (\bibinfo {year} {2012})}\BibitemShut {NoStop}%
\bibitem [{\citenamefont {Iravani}\ \emph {et~al.}(2018)\citenamefont
  {Iravani}, \citenamefont {Sabzyan}, \citenamefont {Vafaee},\ and\
  \citenamefont {Buzari}}]{Iravani2018}%
  \BibitemOpen
  \bibfield  {author} {\bibinfo {author} {\bibfnamefont {H.}~\bibnamefont
  {Iravani}}, \bibinfo {author} {\bibfnamefont {H.}~\bibnamefont {Sabzyan}},
  \bibinfo {author} {\bibfnamefont {M.}~\bibnamefont {Vafaee}}, \ and\ \bibinfo
  {author} {\bibfnamefont {B.}~\bibnamefont {Buzari}},\ }\href {\doibase
  10.1088/1361-6455/aaafb1} {\bibfield  {journal} {\bibinfo  {journal} {Journal
  of Physics B: Atomic, Molecular and Optical Physics}\ }\textbf {\bibinfo
  {volume} {51}},\ \bibinfo {pages} {74003} (\bibinfo {year}
  {2018})}\BibitemShut {NoStop}%
\bibitem [{\citenamefont {Symonds}\ \emph {et~al.}(2015)\citenamefont
  {Symonds}, \citenamefont {Wu}, \citenamefont {Ronto}, \citenamefont {Zagoya},
  \citenamefont {{Figueira De Morisson Faria}},\ and\ \citenamefont
  {Shalashilin}}]{Symonds2015}%
  \BibitemOpen
  \bibfield  {author} {\bibinfo {author} {\bibfnamefont {C.}~\bibnamefont
  {Symonds}}, \bibinfo {author} {\bibfnamefont {J.}~\bibnamefont {Wu}},
  \bibinfo {author} {\bibfnamefont {M.}~\bibnamefont {Ronto}}, \bibinfo
  {author} {\bibfnamefont {C.}~\bibnamefont {Zagoya}}, \bibinfo {author}
  {\bibfnamefont {C.}~\bibnamefont {{Figueira De Morisson Faria}}}, \ and\
  \bibinfo {author} {\bibfnamefont {D.~V.}\ \bibnamefont {Shalashilin}},\
  }\href {\doibase 10.1103/PhysRevA.91.023427} {\bibfield  {journal} {\bibinfo
  {journal} {Physical Review A - Atomic, Molecular, and Optical Physics}\
  }\textbf {\bibinfo {volume} {91}},\ \bibinfo {pages} {1} (\bibinfo {year}
  {2015})}\BibitemShut {NoStop}%
\bibitem [{\citenamefont {Scrinzi}(2012)}]{Scrinzi2012}%
  \BibitemOpen
  \bibfield  {author} {\bibinfo {author} {\bibfnamefont {A.}~\bibnamefont
  {Scrinzi}},\ }\href {\doibase 10.1088/1367-2630/14/8/085008} {\bibfield
  {journal} {\bibinfo  {journal} {New Journal of Physics}\ }\textbf {\bibinfo
  {volume} {14}},\ \bibinfo {pages} {085008} (\bibinfo {year}
  {2012})}\BibitemShut {NoStop}%
\bibitem [{\citenamefont {Bandrauk}\ \emph {et~al.}(2008)\citenamefont
  {Bandrauk}, \citenamefont {Chelkowski}, \citenamefont {Kawai},\ and\
  \citenamefont {Lu}}]{Bandrauk2008}%
  \BibitemOpen
  \bibfield  {author} {\bibinfo {author} {\bibfnamefont {A.~D.}\ \bibnamefont
  {Bandrauk}}, \bibinfo {author} {\bibfnamefont {S.}~\bibnamefont
  {Chelkowski}}, \bibinfo {author} {\bibfnamefont {S.}~\bibnamefont {Kawai}}, \
  and\ \bibinfo {author} {\bibfnamefont {H.}~\bibnamefont {Lu}},\ }\href
  {\doibase 10.1103/PhysRevLett.101.153901} {\bibfield  {journal} {\bibinfo
  {journal} {Physical Review Letters}\ }\textbf {\bibinfo {volume} {101}}
  (\bibinfo {year} {2008}),\ 10.1103/PhysRevLett.101.153901}\BibitemShut
  {NoStop}%
\bibitem [{\citenamefont {Guan}\ \emph {et~al.}(2006)\citenamefont {Guan},
  \citenamefont {Tong},\ and\ \citenamefont {Chu}}]{Guan2006}%
  \BibitemOpen
  \bibfield  {author} {\bibinfo {author} {\bibfnamefont {X.}~\bibnamefont
  {Guan}}, \bibinfo {author} {\bibfnamefont {X.~M.}\ \bibnamefont {Tong}}, \
  and\ \bibinfo {author} {\bibfnamefont {S.~I.}\ \bibnamefont {Chu}},\ }\href
  {\doibase 10.1103/PhysRevA.73.023403} {\bibfield  {journal} {\bibinfo
  {journal} {Physical Review A - Atomic, Molecular, and Optical Physics}\
  }\textbf {\bibinfo {volume} {73}} (\bibinfo {year} {2006}),\
  10.1103/PhysRevA.73.023403}\BibitemShut {NoStop}%
\bibitem [{\citenamefont {Itatani}\ \emph {et~al.}(2004)\citenamefont
  {Itatani}, \citenamefont {Lavesque}, \citenamefont {Zeidler}, \citenamefont
  {Niikura}, \citenamefont {P{\'{e}}pin}, \citenamefont {Kieffer},
  \citenamefont {Corkum},\ and\ \citenamefont {Villeneuve}}]{Itatani2004}%
  \BibitemOpen
  \bibfield  {author} {\bibinfo {author} {\bibfnamefont {J.}~\bibnamefont
  {Itatani}}, \bibinfo {author} {\bibfnamefont {J.}~\bibnamefont {Lavesque}},
  \bibinfo {author} {\bibfnamefont {D.}~\bibnamefont {Zeidler}}, \bibinfo
  {author} {\bibfnamefont {H.}~\bibnamefont {Niikura}}, \bibinfo {author}
  {\bibfnamefont {H.}~\bibnamefont {P{\'{e}}pin}}, \bibinfo {author}
  {\bibfnamefont {J.~C.}\ \bibnamefont {Kieffer}}, \bibinfo {author}
  {\bibfnamefont {P.~B.}\ \bibnamefont {Corkum}}, \ and\ \bibinfo {author}
  {\bibfnamefont {D.~M.}\ \bibnamefont {Villeneuve}},\ }\href {\doibase
  10.1038/nature03183} {\bibfield  {journal} {\bibinfo  {journal} {Nature}\
  }\textbf {\bibinfo {volume} {432}},\ \bibinfo {pages} {867} (\bibinfo {year}
  {2004})}\BibitemShut {NoStop}%
\bibitem [{\citenamefont {Eidi}\ \emph
  {et~al.}(2018{\natexlab{a}})\citenamefont {Eidi}, \citenamefont {Vafaee},\
  and\ \citenamefont {Landsman}}]{Eidi2018b}%
  \BibitemOpen
  \bibfield  {author} {\bibinfo {author} {\bibfnamefont {M.}~\bibnamefont
  {Eidi}}, \bibinfo {author} {\bibfnamefont {M.}~\bibnamefont {Vafaee}}, \ and\
  \bibinfo {author} {\bibfnamefont {A.}~\bibnamefont {Landsman}},\ }\href
  {\doibase 10.3390/app8081252} {\bibfield  {journal} {\bibinfo  {journal}
  {Applied Sciences}\ }\textbf {\bibinfo {volume} {8}},\ \bibinfo {pages}
  {1252} (\bibinfo {year} {2018}{\natexlab{a}})}\BibitemShut {NoStop}%
\bibitem [{\citenamefont {Shalashilin}\ and\ \citenamefont
  {Child}(2004)}]{Shalashilin2004c}%
  \BibitemOpen
  \bibfield  {author} {\bibinfo {author} {\bibfnamefont {D.~V.}\ \bibnamefont
  {Shalashilin}}\ and\ \bibinfo {author} {\bibfnamefont {M.~S.}\ \bibnamefont
  {Child}},\ }\href {\doibase 10.1016/j.chemphys.2004.06.013} {\bibfield
  {journal} {\bibinfo  {journal} {Chemical Physics}\ }\textbf {\bibinfo
  {volume} {304}},\ \bibinfo {pages} {103} (\bibinfo {year}
  {2004})}\BibitemShut {NoStop}%
\bibitem [{\citenamefont {Shalashilin}\ \emph {et~al.}(2008)\citenamefont
  {Shalashilin}, \citenamefont {Child},\ and\ \citenamefont
  {Kirrander}}]{Shalashilin2008a}%
  \BibitemOpen
  \bibfield  {author} {\bibinfo {author} {\bibfnamefont {D.~V.}\ \bibnamefont
  {Shalashilin}}, \bibinfo {author} {\bibfnamefont {M.~S.}\ \bibnamefont
  {Child}}, \ and\ \bibinfo {author} {\bibfnamefont {A.}~\bibnamefont
  {Kirrander}},\ }\href {\doibase 10.1016/j.chemphys.2007.11.006} {\bibfield
  {journal} {\bibinfo  {journal} {Chemical Physics}\ }\textbf {\bibinfo
  {volume} {347}},\ \bibinfo {pages} {257} (\bibinfo {year}
  {2008})}\BibitemShut {NoStop}%
\bibitem [{\citenamefont {Kirrander}\ and\ \citenamefont
  {Shalashilin}(2011)}]{Kirrander2011}%
  \BibitemOpen
  \bibfield  {author} {\bibinfo {author} {\bibfnamefont {A.}~\bibnamefont
  {Kirrander}}\ and\ \bibinfo {author} {\bibfnamefont {D.~V.}\ \bibnamefont
  {Shalashilin}},\ }\href {\doibase 10.1103/PhysRevA.84.033406} {\bibfield
  {journal} {\bibinfo  {journal} {Physical Review A - Atomic, Molecular, and
  Optical Physics}\ }\textbf {\bibinfo {volume} {84}},\ \bibinfo {pages}
  {33406} (\bibinfo {year} {2011})}\BibitemShut {NoStop}%
\bibitem [{\citenamefont {Zhou}\ and\ \citenamefont
  {Chu}(2011{\natexlab{b}})}]{Zhou2011}%
  \BibitemOpen
  \bibfield  {author} {\bibinfo {author} {\bibfnamefont {Z.}~\bibnamefont
  {Zhou}}\ and\ \bibinfo {author} {\bibfnamefont {S.~I.}\ \bibnamefont {Chu}},\
  }\href {\doibase 10.1103/PhysRevA.83.033406} {\bibfield  {journal} {\bibinfo
  {journal} {Physical Review A - Atomic, Molecular, and Optical Physics}\
  }\textbf {\bibinfo {volume} {83}},\ \bibinfo {pages} {33406} (\bibinfo {year}
  {2011}{\natexlab{b}})}\BibitemShut {NoStop}%
\bibitem [{\citenamefont {Eidi}\ \emph {et~al.}(2016)\citenamefont {Eidi},
  \citenamefont {Vafaee}, \citenamefont {Niknam},\ and\ \citenamefont
  {Morshedian}}]{Eidi2016}%
  \BibitemOpen
  \bibfield  {author} {\bibinfo {author} {\bibfnamefont {M.}~\bibnamefont
  {Eidi}}, \bibinfo {author} {\bibfnamefont {M.}~\bibnamefont {Vafaee}},
  \bibinfo {author} {\bibfnamefont {A.~R.}\ \bibnamefont {Niknam}}, \ and\
  \bibinfo {author} {\bibfnamefont {N.}~\bibnamefont {Morshedian}},\ }\href
  {\doibase 10.1016/j.cplett.2016.04.054} {\bibfield  {journal} {\bibinfo
  {journal} {Chemical Physics Letters}\ }\textbf {\bibinfo {volume} {653}},\
  \bibinfo {pages} {60} (\bibinfo {year} {2016})}\BibitemShut {NoStop}%
\bibitem [{\citenamefont {Eidi}\ \emph
  {et~al.}(2018{\natexlab{b}})\citenamefont {Eidi}, \citenamefont {Vafaee},\
  and\ \citenamefont {Rooein}}]{Eidi2018a}%
  \BibitemOpen
  \bibfield  {author} {\bibinfo {author} {\bibfnamefont {M.}~\bibnamefont
  {Eidi}}, \bibinfo {author} {\bibfnamefont {M.}~\bibnamefont {Vafaee}}, \ and\
  \bibinfo {author} {\bibfnamefont {M.}~\bibnamefont {Rooein}},\ }\href
  {\doibase 10.1002/jcc.25133} {\bibfield  {journal} {\bibinfo  {journal}
  {Journal of Computational Chemistry}\ }\textbf {\bibinfo {volume} {39}},\
  \bibinfo {pages} {679} (\bibinfo {year} {2018}{\natexlab{b}})}\BibitemShut
  {NoStop}%
\bibitem [{\citenamefont {Schultz}\ and\ \citenamefont
  {Vrakking}(2014)}]{Schultz2014}%
  \BibitemOpen
  \bibfield  {author} {\bibinfo {author} {\bibfnamefont {T.}~\bibnamefont
  {Schultz}}\ and\ \bibinfo {author} {\bibfnamefont {M.}~\bibnamefont
  {Vrakking}},\ }\href {\doibase 10.1002/9783527677689} {\emph {\bibinfo
  {title} {Attosecond and XUV Physics: Ultrafast Dynamics and Spectroscopy}}},\
  Vol.\ \bibinfo {volume} {9783527411}\ (\bibinfo  {publisher} {Wiley
  Blackwell},\ \bibinfo {year} {2014})\ pp.\ \bibinfo {pages}
  {1--607}\BibitemShut {NoStop}%
\bibitem [{\citenamefont {Bandrauk}\ and\ \citenamefont
  {Shen}(1993)}]{Bandrauk1993}%
  \BibitemOpen
  \bibfield  {author} {\bibinfo {author} {\bibfnamefont {A.~D.}\ \bibnamefont
  {Bandrauk}}\ and\ \bibinfo {author} {\bibfnamefont {H.}~\bibnamefont
  {Shen}},\ }\href {\doibase 10.1063/1.465362} {\bibfield  {journal} {\bibinfo
  {journal} {The Journal of Chemical Physics}\ }\textbf {\bibinfo {volume}
  {99}},\ \bibinfo {pages} {1185} (\bibinfo {year} {1993})}\BibitemShut
  {NoStop}%
\bibitem [{\citenamefont {Ahmadi}\ \emph {et~al.}(2014)\citenamefont {Ahmadi},
  \citenamefont {Maghari}, \citenamefont {Sabzyan}, \citenamefont {Niknam},\
  and\ \citenamefont {Vafaee}}]{Ahmadi2014}%
  \BibitemOpen
  \bibfield  {author} {\bibinfo {author} {\bibfnamefont {H.}~\bibnamefont
  {Ahmadi}}, \bibinfo {author} {\bibfnamefont {A.}~\bibnamefont {Maghari}},
  \bibinfo {author} {\bibfnamefont {H.}~\bibnamefont {Sabzyan}}, \bibinfo
  {author} {\bibfnamefont {A.~R.}\ \bibnamefont {Niknam}}, \ and\ \bibinfo
  {author} {\bibfnamefont {M.}~\bibnamefont {Vafaee}},\ }\href {\doibase
  10.1103/PhysRevA.90.043411} {\bibfield  {journal} {\bibinfo  {journal}
  {Physical Review A - Atomic, Molecular, and Optical Physics}\ }\textbf
  {\bibinfo {volume} {90}},\ \bibinfo {pages} {43411} (\bibinfo {year}
  {2014})}\BibitemShut {NoStop}%
\bibitem [{\citenamefont {Vafaee}\ and\ \citenamefont
  {Sabzyan}(2004)}]{Vafaee2004}%
  \BibitemOpen
  \bibfield  {author} {\bibinfo {author} {\bibfnamefont {M.}~\bibnamefont
  {Vafaee}}\ and\ \bibinfo {author} {\bibfnamefont {H.}~\bibnamefont
  {Sabzyan}},\ }\href {\doibase 10.1088/0953-4075/37/20/009} {\bibfield
  {journal} {\bibinfo  {journal} {Journal of Physics B: Atomic, Molecular and
  Optical Physics}\ }\textbf {\bibinfo {volume} {37}},\ \bibinfo {pages} {4143}
  (\bibinfo {year} {2004})}\BibitemShut {NoStop}%
\bibitem [{\citenamefont {Corkum}\ \emph {et~al.}(1994)\citenamefont {Corkum},
  \citenamefont {Burnett},\ and\ \citenamefont {Ivanov}}]{Corkum1994}%
  \BibitemOpen
  \bibfield  {author} {\bibinfo {author} {\bibfnamefont {P.~B.}\ \bibnamefont
  {Corkum}}, \bibinfo {author} {\bibfnamefont {N.~H.}\ \bibnamefont {Burnett}},
  \ and\ \bibinfo {author} {\bibfnamefont {M.~Y.}\ \bibnamefont {Ivanov}},\
  }\href {\doibase 10.1364/ol.19.001870} {\bibfield  {journal} {\bibinfo
  {journal} {Optics Letters}\ }\textbf {\bibinfo {volume} {19}},\ \bibinfo
  {pages} {1870} (\bibinfo {year} {1994})}\BibitemShut {NoStop}%
\bibitem [{\citenamefont {Krause}\ \emph
  {et~al.}(1992{\natexlab{b}})\citenamefont {Krause}, \citenamefont {Schafer},\
  and\ \citenamefont {Kulander}}]{Krause1992b}%
  \BibitemOpen
  \bibfield  {author} {\bibinfo {author} {\bibfnamefont {J.~L.}\ \bibnamefont
  {Krause}}, \bibinfo {author} {\bibfnamefont {K.~J.}\ \bibnamefont {Schafer}},
  \ and\ \bibinfo {author} {\bibfnamefont {K.~C.}\ \bibnamefont {Kulander}},\
  }\href {\doibase 10.1103/PhysRevA.45.4998} {\bibfield  {journal} {\bibinfo
  {journal} {Physical Review A}\ }\textbf {\bibinfo {volume} {45}},\ \bibinfo
  {pages} {4998} (\bibinfo {year} {1992}{\natexlab{b}})}\BibitemShut {NoStop}%
\bibitem [{\citenamefont {Helgaker}\ and\ \citenamefont
  {Taylor}(1995)}]{Helgaker1995}%
  \BibitemOpen
  \bibfield  {author} {\bibinfo {author} {\bibfnamefont {T.}~\bibnamefont
  {Helgaker}}\ and\ \bibinfo {author} {\bibfnamefont {P.~R.}\ \bibnamefont
  {Taylor}}\ }(\bibinfo {year} {1995})\ pp.\ \bibinfo {pages}
  {725--856}\BibitemShut {NoStop}%
\bibitem [{\citenamefont {Joe}\ and\ \citenamefont {Kuo}(2003)}]{Joe2003}%
  \BibitemOpen
  \bibfield  {author} {\bibinfo {author} {\bibfnamefont {S.}~\bibnamefont
  {Joe}}\ and\ \bibinfo {author} {\bibfnamefont {F.~Y.}\ \bibnamefont {Kuo}},\
  }\href {\doibase 10.1145/641876.641879} {\bibfield  {journal} {\bibinfo
  {journal} {ACM Transactions on Mathematical Software}\ }\textbf {\bibinfo
  {volume} {29}},\ \bibinfo {pages} {49} (\bibinfo {year} {2003})}\BibitemShut
  {NoStop}%
\bibitem [{\citenamefont {Safaei}(2018)}]{Safaei2018}%
  \BibitemOpen
  \bibfield  {author} {\bibinfo {author} {\bibfnamefont {N.}~\bibnamefont
  {Safaei}},\ }\href {\doibase 10.1088/1612-202X/aa9318} {\bibfield  {journal}
  {\bibinfo  {journal} {Laser Physics Letters}\ }\textbf {\bibinfo {volume}
  {15}},\ \bibinfo {pages} {15202} (\bibinfo {year} {2018})}\BibitemShut
  {NoStop}%
\end{thebibliography}%

\end{document}